\documentclass[journal]{IEEEtran}
\usepackage{amsmath,amsfonts}
\usepackage{algorithmic}
\usepackage{algorithm}
\usepackage{array}
\usepackage[caption=false,font=normalsize,labelfont=sf,textfont=sf]{subfig}
\usepackage{textcomp}
\usepackage{stfloats}
\usepackage{url}
\usepackage{verbatim}
\usepackage{graphicx}
\usepackage{cite}
\usepackage{color}
\begin{document}

\title{Computation Offloading for Edge Computing in RIS-Assisted Symbiotic Radio Systems}

\author{Bin Li, Zhen Qian, Lei Liu, \IEEEmembership{Member, IEEE}, Yuan Wu, \IEEEmembership{Senior Member, IEEE}, Dapeng Lan, \IEEEmembership{Member, IEEE},\\
	 and Celimuge Wu, \IEEEmembership{Senior Member, IEEE}
        % <-this % stops a space

\thanks{Bin Li and Zhen Qian are with the School of Computer Science, Nanjing University of Information Science and Technology, Nanjing, 210044, China (e-mail: bin.li@nuist.edu.cn; zhenqian@nuist.edu.cn).}
\thanks{Lei Liu is with the Xidian Guangzhou Institute of Technology, Guangzhou 510555, China (e-mail: tianjiaoliulei@163.com).}
\thanks{Yuan Wu is with the State Key Laboratory of Internet of Things for Smart City, University of Macau, Macao, China, and also with the Department of Computer and Information Science, University of Macau (email: yuanwu@um.edu.mo).}
\thanks{Dapeng Lan is with Techforgood AS, Norway, and also with University of Oslo, Norway (e-mail: dapengl@uio.no).}
\thanks{Celimuge Wu is with Meta-Networking Research Center, The University of Electro-Communications, 1-5-1, Chofugaoka, Chofu-shi, Tokyo,182-8585 Japan (e-mail: celimuge@uec.ac.jp).}
}

% The paper headers
%\markboth{Journal of \LaTeX\ Class Files,~Vol.~XX, No.~X, XXXXX~XXXX}%
%{Shell \MakeLowercase{\textit{et al.}}: A Sample Article Using IEEEtran.cls for IEEE Journals}

% \IEEEpubid{0000--0000/00\$00.00~\copyright~2021 IEEE}
% Remember, if you use this you must call \IEEEpubidadjcol in the second
% column for its text to clear the IEEEpubid mark.

\maketitle

\begin{abstract}
  In the paper, we investigate the coordination process of sensing and computation offloading in a reconfigurable intelligent surface (RIS)-aided base station (BS)-centric symbiotic radio (SR) systems. Specifically, the Internet-of-Things (IoT) devices first sense data from environment and then tackle the data locally or offload the data to BS for remote computing, while RISs are leveraged to enhance the quality of blocked channels and also act as IoT devices to transmit its sensed data. To explore the mechanism of cooperative sensing and computation offloading in this system, we aim at maximizing the total completed sensed bits of all users and RISs by jointly optimizing the time allocation parameter, the passive beamforming at each RIS, the transmit beamforming at BS, and the energy partition parameters for all users subject to the size of sensed data, energy supply and given time cycle. The formulated nonconvex problem is tightly coupled by the time allocation parameter and involves the mathematical expectations, which cannot be solved straightly. We use Monte Carlo and fractional programming methods to transform the nonconvex objective function and then propose an alternating optimization-based algorithm to find an approximate solution with guaranteed convergence. Numerical results show that the RIS-aided SR system outperforms other benchmarks in sensing. Furthermore, with the aid of RIS, the channel and system performance can be significantly improved.
\end{abstract}


\section{Introduction}
\IEEEPARstart{T}{he} large-scale commercialization of fifth-generation (5G) has promoted the network to enter a new era of Internet of Things (IoT), and simultaneously accompanies with billions of IoT devices with sensing, computing, and wireless communication capabilities \cite{1}. Particularly, 5G high-speed communication and massive sensor nodes have spawned various novel applications for users, involving resource-intensive and latency-sensitive applications such as intelligent transportation and virtual reality \cite{DuTNSE2022,addtwotwo,addtwothree}. However, these novel applications require not only fresh data, but also a large amount of hashing power. Generally, sensor nodes send the sensed data to the central cloud server, and then users request it from the central cloud server, which may render the data obsolete. Additionally, massive sensor nodes will lead to a shortage of spectrum resources and low-rate communication. Thus, novel applications significantly increase the burden of computation and communication on users.

In order to tackle the challenges posed by diversified emerging applications, improving communication and computation efficiencies are two effective means to alleviate the pressure on users. For one thing, symbiotic radio (SR) system \cite{3}, which is developed from the backscatter technique \cite{3-1,3-2,3-3}, has recently been recognized as a novel cooperative communication system for enabling both passive communication of the secondary system and active communication of the primary system at the same time. Specifically, the secondary system utilizes the incident signals from the primary system to enable its own wireless communication without any dedicated radio-frequency (RF) components, and in return provides the multi-path channels for the primary system. Consequently, SR-based system can achieve mutual benefit and high spectral efficiency among massive IoT devices. However, compared to the primary system, the rate performance of the conventional secondary system has a certain gap because of the double fading channels. Meanwhile, reconfigurable intelligent surface (RIS)-aided communication \cite{4,5}, which is another promising communication paradigm, can improve wireless communication, such as enhancing blocked-channel communication and enlarging the coverage of the base station (BS). It compensates for the shortcomings of the secondary system in SR systems. Therefore, RIS-assisted SR system is envisioned as an enabling technology to realize large-scale IoT devices connectivity. For another, mobile edge computing (MEC) \cite{Dujianbo2020,6,7} is able to provide the users with low-latency and high-performance computing services at the network edge integrated with servers. Thus, massive IoT devices can relieve from their heavy workloads by taking full use of limited resources with the help of MEC.

Recently, lots of works have shown that the combination of RIS and MEC can improve energy efficiency, computation efficiency and rate performance \cite{8,9,10,11,12,13}. Particularly, the authors of \cite{8} studied an RIS-assisted multiuser MEC network, where users’ energy efficiency and computation efficiency are maximized through the proposed deep learning-based architecture. Later in \cite{9}, a delay-oriented resource allocation algorithm was proposed to exploit the spare computing resources of IoT devices for RIS-aided device-to-device cooperative computing architectures. To increase the task offloading rate, the authors of \cite{10} investigated the sum delay problem for RIS-aided MEC systems. In \cite{11}, the latency minimization problem was studied to evaluate the benefits of using RIS in the MEC system. The authors of \cite{12} explored the RIS benefits on the overall computing performance of the MEC system by maximizing the sum computational bits. In \cite{13}, the advantages of RIS were utilized to improve the efficiency of wireless powered and task offloading, where the total computation bits maximization problem is solved with an iterative algorithm (AO).

For RIS-assisted MEC systems, RIS usually acts as an auxiliary device to enhance the channels quality of IoT devices. Although RIS is able to enhance the desired signals, it also inevitably strengthens the noise signals. In the meanwhile, massive IoT devices will contribute to a shortage of system resources, e.g., spectrum and energy resources. Providentially, SR-based systems have attracted much attention for its spectrum- and energy-efficiency, providing a new idea for mitigating these drawbacks. In view of the benefits of SR, recent efforts have been conducted to investigate the applications of SR techniques, especially in rate efficiency and RIS-aided systems. To be specific, in \cite{3}, the maximization problem of weighted sum-rate (WSR) and the minimization problem of transmitting power were first studied in two basic SR systems. The authors of \cite{14} investigated a full-duplex-enabled SR network, where the primary transmitter can transmit primary signals and receive backscatter signals simultaneously through full-duplex communications. For the applications of SR model in RIS-aided system, the RIS passive beamforming and BS active beamforming are jointly optimized in \cite{15} to minimize the primary transmit power to show the benefits of introducing RIS in the SR system. The authors of \cite{16} considered the downlink transmission scenario with multiple users and RISs, where each RIS assists its nearby users to improve the channel quality and simultaneously acts as an IoT device to transmit its sensed data to the users. This work was extended in \cite{17} to the case of uplink scenario, where the WSR of all RISs and users is maximized by jointly optimizing passive and active beamforming.

However, to our best knowledge, all of the above works are not further investigated the process of combining sensing and computation offloading. In the forthcoming B5G era, sensing data has a wide range of applications, such as smart transportation and environment monitoring, which is of high practical significance. To further exploit the internal mechanism of combing sensing and computation offloading, there have already been some works that explore the integrated data sensing and computation offloading in MEC architecture \cite{18,19,20,addtwentyone}. Particularly, the work in \cite{18} considered a “\emph{sense-then-offload}” MEC system, where users’ offloading throughput of sensed data is maximized under two communication protocols. The authors of \cite{19} investigated the coordination of unmanned aerial vehicles sensing and computation, where the overall system utility is maximized by a deep reinforcement learning method. In \cite{20}, the authors constructed the sensing, communication, and computation networks, and studied the device association and subchannel assignment problems via matching algorithm. In \cite{addtwentyone}, the authors leveraged RIS to enhance MEC performance in the integrated sensing and communication.

Sparked by the above discussions, this paper considers the data sensing and task offloading in an RIS-aided BS-centric SR system. Note that unlike the existing works on WSR \cite{15,16}, the “\emph{sense-then-offload}” protocol is adopted in this paper to further study the internal mechanism of sensing and computation offloading. In the SR system, each RIS can modulate its own information through the incident signals from a nearby user, which allows RIS to transmit its sensed data to BS without any RF components and improve the spectrum and energy efficiencies of the system. Besides, we assume that each RIS is equipped with micro-sensors, which can sense data from the surrounding environment, such as noise level and temperature. Under this system, such current environment information can be first collected by RISs and eventually aggregated at BS. The results of tackling sensed data can provide convenience to the users near RIS in return.

The main contributions of this work are summarized as follows.

\begin{itemize}
  \item We investigate the coordination process of sensing and computation offloading in the RIS-aided BS-centric SR system. To trade off the performance between sensing and computation offloading, we formulate the total completed sensed task maximization problem of all users and RISs by jointly optimizing the time allocation parameter, the passive beamforming at each RIS, the transmit beamforming at BS, and the energy partition parameters for all users under the constraints on the size of sensed data, energy supply and given time cycle.
  \item The formulated problem is tightly coupled by the time allocation parameter and involves the mathematical expectations, which is challenging to find the global optimal solution. To address the nonconvex problem, we first tackle the formulated problem effectively under different given time allocation parameters through both the fractional programming (FP) and AO techniques with guaranteed convergence. Then, the approximate time allocation parameter can be obtained by the results of the first step.
  \item Via simulation results, we show that the RIS-aided SR system outperforms other benchmarks in sensing. Besides, with the aid of RIS, not only the channel is enhanced, but also the total completed sensed data is significantly improved.
\end{itemize}

The remainder of this paper is organized as follows. Section II describes the system model and formulates the optimization problem. Section III proposes an AO-based algorithm to solve the mathematical formulated problem with guaranteed convergence. The numerical results and conclusions are given in detail in Section IV and Section V, respectively.

\emph{Notations}---In this paper, the matrices and vectors are bold uppercase and bold lowercase, respectively. ${{\left( \cdot  \right)}^{\text{H}}}$, ${{\left( \cdot  \right)}^{\text{*}}}$, $\angle \left( \cdot  \right)$, $\text{Re}\left\{ \cdot  \right\}$, and $\mathbb{E}\left\{ \cdot  \right\}$ denote conjugate transpose, conjugation, phase, real dimension, and expectation, respectively. $\mathcal{C}\mathcal{N}\left( \mu ,{{\sigma }^{2}} \right)$ denotes the distribution of a circularly symmetric complex Gaussian. $\text{tr}\left( \mathbf{A} \right)$ denotes the trace of square matrix $\mathbf{A}$. Additionally, $\partial f\left( x \right)/\partial x$ and ${{\partial }^{2}}f\left( x \right)/\partial {{x}^{2}}$ denote the first and second derivative of the function $f\left( x \right)$ with respect to $x$, respectively.

\begin{figure}[!t]
  \centering
  \includegraphics[width=3in]{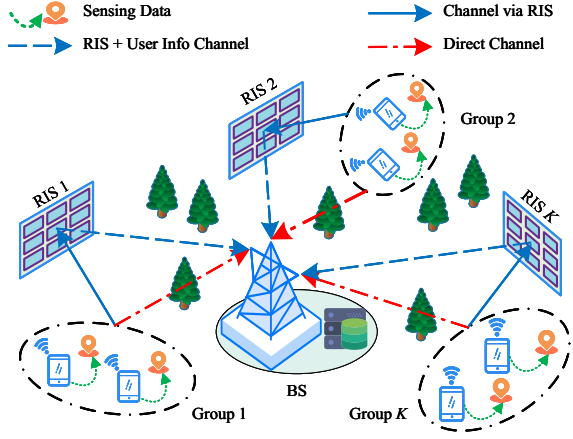}
  \caption{Illustration of the RIS-aided BS-centric SR system.}
  \label{fig_1}
\end{figure}
\section{System Model And Problem Formulation}
We consider an RIS-aided BS-centric SR system as depicted in Fig. 1, which consists of $K$ RIS-aided user groups and one $M$-antenna BS integrated with an edge server. Specifically, each user group includes multiple single-antenna users and is equipped with one RIS with $N$ reflecting elements. By incorporating an SR-based system, RISs can not only improve the quality of the blocked channels between users and BS, but also act as IoT devices that use backscattering technology to transmit their sensed data to BS simultaneously. To be more specific, every RIS modulates its information through the incident signals from nearby users and reflects it to BS. This considered system allows RISs to send their own information to BS without any dedicated RF source and senses data from the surrounding environment, which is of practical interest, e.g., enhancing the blocked channel and providing the more IoT device connectivity for smart cities. To simplify the system model, we consider that each user group contains only one user, and other users in the same user group can obtain service from the nearby RIS in a time division multiple access mode. For ease of expression, we define $k\in \mathcal{K}=\{1,2,\cdots ,K\}$ as the set of all users, and $n\in \mathcal{N}=\left\{ 1,2,\cdots ,N \right\}$ as the set of reflecting elements for each RIS, respectively. Additionally, we assume that the nearby RIS to user $k$ is denoted as RIS $k$. Thus, user $k$ transmits sensed data to BS, and simultaneously enables the RIS $k$ to transmit its sensed data to BS. Moreover, we consider that multiple RISs and user groups are far apart so that the signal of user $k$ is only reflected by the nearby RIS (i.e., RIS $k$) and the reflected signals from the far RISs are ignored. Thus, for the links associated with RIS, we only consider the link between user $k$ and RIS $k$ in this paper.

\subsection{Sensing Model}
By following \cite{18}, a “\emph{sense-then-offload}” mode is considered in this paper, in which each IoT device first senses data from the target region and then addresses the sensed data. We first introduce a ratio parameter $\alpha \in \left[ 0,1 \right]$ for dividing the sensing time and addressing time during a given time slot $T$, and $\left( 1-\alpha  \right)T$ seconds will be used for sensing data from the environment while $\alpha T$ seconds will be used for processing sensed data. In reality, a variety of sensors, including cameras and microphones \cite{18-2}, are leveraged in different sensing applications, collecting sensing data with varying sample rates, where sample rate is the number of samples obtained in one second \cite{18-1}. Thus, the rate of sensing data can be computed as the size of samples times the sample rate, and is related to the kinds of environment data. For sensing temperature, the sensing rate is small. For sensing images, the sensing rate is large. Let $v_{k}^{\text{p}}$ and $v_{k}^{\text{s}}$ denote the sensing rate of user $k$ and RIS $k$, respectively. In this case, the data sensed by user $k$ and RIS $k$ during the sensing process can be respectively expressed as
\begin{align}
  \label{eq1}
  M_{k}^{\text{s}}=(1-\alpha )Tv_{k}^{\text{s}},\\
  M_{k}^{\text{p}}=(1-\alpha )Tv_{k}^{\text{p}}.
\end{align}

During the sensing process, we assume that RISs are embedded with some environmental sensors, such as noise level, temperature, etc., which can provide the sensing capability of RISs. When RISs transmit the sensed data to BS, RISs can achieve low-power communication with the help of user’s signal and backscatter technique \cite{21} in the SR system. Such sensors embedded on the RISs are usually powered by green energy such as light and wind, and the backscatter-based communication utilizes the user’s signal energy, so the energy consumed by the RISs during the whole process is ignored. In addition, each user uses its own battery energy for sensing, and the corresponding energy consumption of user $k$ during the sensing process is represented by
\begin{equation}
  \label{eq2}
  E_k^{\rm{s}} = p_k^{\rm{s}}M_k^{\rm{p}},
\end{equation}
where $p_{k}^{s}$ is the energy cost per unit for data sensing.

\subsection{Symbiotic Radio Communication Model}
In this paper, we adopt the block flat-fading channel. The channels from user $k$ to RIS $k$, from RIS $k$ to BS, from user $k$ to BS, are denoted by ${\mathbf{h}_{\text{r},k}}\in {{\mathbb{C}}^{N\times 1}}$, ${\mathbf{G}_{k}}\in {{\mathbb{C}}^{N\times M}}$, and ${\mathbf{h}_{\text{d},k}}\in {{\mathbb{C}}^{M\times 1}}$, respectively. Particularly, the channels between users and BS are blocked by obstacles and extremely weak. Following \cite{16} and \cite{22}, the technologies of channel estimation can be found and thus we assume that the perfect channel state information is available.

Let ${{c}_{k}}$ denote the signal transmitted by RIS $k$ to BS, and we assume that ${{c}_{k}}$ adopts \emph{binary phase-shift keying} modulation, i.e., ${{c}_{k}}=\left\{ -1,1 \right\}$. Moreover, as in \cite{3}, we assume that every time user $k$ transmits $Q$ symbols to BS, an IoT symbol is transmitted from RIS $k$, i.e., $Q{{T}_\text{p}}={{T}_\text{s}}$, where $Q$ is much greater than 1, ${{T}_\text{s}}$ is the secondary symbol period from the RIS, and ${{T}_\text{p}}$ is the primary symbol period from the user. Accordingly, the phase shift matrix of RIS $k$ can be expressed as ${{\mathbf{\Theta }}_{k}}={{\zeta }_{k}}\text{diag}\left( \varphi _{1}^{k},\varphi _{2}^{k},\cdots ,\varphi _{N}^{k} \right)\in {{\mathbb{C}}^{N\times N}}$, where $\varphi _{n}^{k}={{\text{e}}^{\text{j}\theta _{n}^{k}}}$ is the phase shift of the $n$-th reflecting element at RIS $k$ with $\theta _{n}^{k}\in \left[ 0,2\pi  \right]$, and ${{\zeta }_{k}}\in \left[ 0,1 \right]$ is the $k$-th RIS reflection amplitude. Note that the joint optimization of phase shift and reflection amplitude of RIS elements can be theoretically analyzed, the amplitude adjustment of RIS reflection is difficult to implement with practical hardware. Thus, by following \cite{27}, we set ${{\zeta }_{k}}=1$ in the sequel of this paper to maximize the reflection efficiency of RIS. In addition, let ${{s}_{k}}\left( q \right)$ denote the $q$-th signal transmitted by user $k$ to BS during one symbol period of ${{c}_{k}}$ and we note that ${{s}_{k}}\left( q \right)\sim\mathcal{C}\mathcal{N}\left( 0,1 \right)$. For notational simplicity, we introduce ${\mathbf{b}_k} = \mathbf{G}_k^{\rm{H}}{\mathbf{\Theta} _k}{\mathbf{h}_{{\rm{r}},k}} \in {\mathbb{C}^{M \times 1}}$ here to express the backscatter channel from user $k$ to BS via RIS $k$. This means when RIS $k$ transmits ${{c}_{k}}=-1$ to BS, it uses $-{{\mathbf{\Theta} }_{k}}$ to reflect its signals. Therefore, we can express the corresponding $q$-th received signals during a period of ${{T}_{s}}$ at BS as
\begin{equation}
  \label{eq3}
  \mathbf{y}\left( q \right)=\sum\limits_{k=1}^{K}{\sqrt{{{p}_{k}}}\left( {\mathbf{h}_{\text{d},k}}+{{c}_{k}}{\mathbf{b}_{k}} \right){{s}_{k}}\left( q \right)}+\mathbf{n},
\end{equation}
where $\mathbf{n}=\left[ {{n}_{1}},{{n}_{2}},\cdots ,{{n}_{M}} \right]\in {{\mathbb{C}}^{M\times 1}}$, ${{n}_{m}}\sim \mathcal{C}\mathcal{N}\left( 0,{{\sigma }^{2}} \right)$ is the noise at the $m$-th antenna of BS, and ${{p}_{k}}$ is the transmit power of user $k$.

After receiving signals, the BS decodes the signals ${{s}_{k}}\left( q \right)$ and ${{c}_{k}}$ with successive interference cancellation (SIC) technique \cite{17}. Specifically, this method is based on the priority of the channel condition to decode signals, where ${{\left| \sqrt{{{p}_{k}}}{\mathbf{h}_{\text{d},k}} \right|}^{2}}+{{\left| \sqrt{{{p}_{k}}}{\mathbf{b}_{k}} \right|}^{2}}$ is used to denote the channel condition of user $k$. This means that users with higher signal energy will be prioritized decoding. The BS cancels decoded signals from (\ref{eq3}) and then decodes the next signals.  Without loss of generality, we assume that the users with higher priority have smaller index. To this end, the intermediate signal used to decoding the ${{s}_{k}}\left( q \right)$ and ${{c}_{k}}$ after receiving beamforming can be denoted as
\begin{equation}
  \label{eq4}
  {{y}_{k}}\left( q \right)=\sum\limits_{i=k}^{K}{\sqrt{{{p}_{i}}}\mathbf{w}_{k}^{\text{H}}\left( {\mathbf{h}_{\text{d},i}}+{{c}_{i}}{\mathbf{b}_{i}} \right){{s}_{i}}\left( q \right)}+\mathbf{w}_{k}^{\text{H}}\mathbf{n}.
\end{equation}

It is noted that the BS first decodes the ${{s}_{k}}\left( q \right)$ by treating the ${\mathbf{b}_k}$ as an additional multi-path component to solve the coupled problem of ${{s}_{k}}\left( q \right)$ and ${{c}_{k}}$ in Eq. (\ref{eq4}). As a result, we can obtain the signal-to-interference-plus-noise ratio (SINR) for decoding ${{s}_{k}}\left( q \right)$ as
\begin{equation}
  \label{eq5}
  \gamma _{k}^{\text{p}}\left( {{c}_{k}} \right)=\frac{{{\left| \sqrt{{{p}_{k}}}\mathbf{w}_{k}^{\text{H}}\left( {\mathbf{h}_{\text{d},k}}+{{c}_{k}}{\mathbf{b}_{k}} \right) \right|}^{2}}}{\sum\limits_{i=k+1}^{K}{{{\left| \sqrt{{{p}_{i}}}\mathbf{w}_{k}^{\text{H}}{\mathbf {h}_{\text{d},i}} \right|}^{2}}}+\sum\limits_{i=k+1}^{K}{{{\left| \sqrt{{{p}_{i}}}\mathbf{w}_{k}^{\text{H}}{\mathbf {b}_{i}} \right|}^{2}}}+{\sigma }_{k}^{2}},
\end{equation}
where $\sigma _{k}^{2}=\left\| {\mathbf{w}_{k}^{\text{H}}} \right\|_{2}^{2}{{\sigma }^{2}}$. Since ${{c}_{k}}$ is a random variable, the rate of decoding ${{s}_{k}}\left( q \right)$ cannot be obtained directly. By taking an expectation of ${{c}_{k}}$, the corresponding average rate of decoding ${{s}_{k}}\left( q \right)$ can be approximated as
\begin{equation}
  \label{eq6}
  r_{k}^{\text{p}}\approx B{{\mathbb{E}}_{{{c}_{k}}}}\left\{ {{\log }_{2}}\left( 1+\gamma _{k}^{\text{p}}\left( {{c}_{k}} \right) \right) \right\},
\end{equation}
where $B$ is the system bandwidth.

After finishing decoding ${{s}_{k}}\left( q \right)$, the SIC technique is utilized to remove ${{s}_{k}}\left( q \right)$ related components from received signals, and then the maximal ratio combing technique is applied to decode ${{c}_{k}}$. Thus, according to \cite{15}, the SINR for decoding ${{c}_{k}}$ can be approximated as (assuming $Q\gg 1$)
\begin{equation}
  \label{eq7}
  \gamma _{k}^{\text{s}}\approx\frac{Q{{\left| \sqrt{{{p}_{k}}}\mathbf{w}_{k}^{\text{H}}{\mathbf {b}_{k}} \right|}^{2}}}{\sum\limits_{i=k+1}^{K}{{{\left| \sqrt{{{p}_{i}}}\mathbf{w}_{k}^{\text{H}}{\mathbf {h}_{\text{d},i}} \right|}^{2}}}+\sum\limits_{i=k+1}^{K}{{{\left| \sqrt{{{p}_{i}}}\mathbf{w}_{k}^{\text{H}}{\mathbf {b}_{i}} \right|}^{2}}}+\sigma _{k}^{2}}.
\end{equation}

We consider that the SINR for decoding ${{c}_{k}}$ is increased by $Q$ times in Eq. (\ref{eq7}) and the corresponding rate of decoding symbol ${{c}_{k}}$ is decreased by $1/Q$. In this case, the rate of decoding ${{c}_{k}}$ can be approximated as
\begin{equation}
  \label{eq8}
  r_{k}^{\text{s}}\approx B{Q_{}}^{-1}{{\log }_{2}}\left( 1+\gamma _{k}^{\text{s}} \right).
\end{equation}

\subsection{Computation Model}
At the beginning of cycle $T$, each user $k$ has an energy budget $E_{k}^{\text{max}}$ dedicated to data sensing and offloading. After finishing the sensing process, each user processes the intensive sensed data (e.g., smart transportation and image recognition) with the remaining energy $E_{k}^{\text{o}}=E_{k}^{\text{max}}-E_{k}^{\text{s}}$ during a time slot $\alpha T$. In this paper, a partial offloading mode is adopted to divide users’ sensed data into two parallel portions for local computing at user itself and edge computing at BS. Consequently, the total energy consumption of each user during second phase includes local computation and offloading sensed data. The BS is equipped with a high-performance edge server for assisting users and RISs to handle their offloaded sensed data and has a large energy supply. Furthermore, RISs act as passive sensing devices that cannot process the sensed data by themselves. As such, RISs can only send it to BS through backscattering technology to process the sensed data.

Let ${{\beta }_{k}}\in \left[ 0,1 \right]$ denote the energy partition parameter for user $k$, and $\left( {1 - {\beta _k}} \right)E_k^{\rm{o}}$ will be used for local computing while ${{\beta }_{k}}E_{k}^{\text{o}}$ will be used for computation offloading. Then, the transmit power of user $k$ for offloading can be represented by
\begin{equation}
  \label{eq9}
  {{p}_{k}}\left( {{\beta }_{k}} \right)={{{\beta }_{k}}E_{k}^{\text{o}}}{{(\alpha T)}_{}}^{-1}.
\end{equation}

After calculating the value of transmit power, the sensed data transmitted by user $k$ and RIS $k$ to BS during the $\alpha T$ can be respectively expressed as $r_{k}^{\text{p}}\alpha T$ and $r_{k}^{\text{s}}\alpha T$ according to Eq. (\ref{eq6}) and Eq. (\ref{eq8}). Since RISs can only deal with sensed data in an offloading manner, the total completed sensed data handled by RIS $k$ during the offloading process can be expressed as
\begin{equation}
  \label{eq10}
  R_{k}^{\text{s}}=r_{k}^{\text{s}}\alpha T.
\end{equation}

Unlike RISs, the users can handle the sensed data locally. To improve the energy efficiency of computing, the voltage and frequency scaling technique \cite{8} is adopted. Therefore, the energy consumption of user $k$ for local computation during $\alpha T$ can be represented by $\alpha T{\kappa}_{l} f_{k}^{3}$. Particularly, $\left( 1-{{\beta }_{k}} \right)E_{k}^{\text{o}}$ of energy is used for local computing. To this end, we have the equation $\left( 1-{{\beta }_{k}} \right)E_{k}^{\text{o}}=\alpha T{\kappa}_{l} f_{k}^{3}$, and we can further express the local computing frequency of user $k$ as
\begin{equation}
  \label{eq11}
  {{f}_{k}}=\sqrt[3]{\frac{(1-{{\beta }_{k}})E_{k}^{\text{o}}}{\alpha T{{\kappa }_{l}}}},
\end{equation}
where ${{\kappa }_{l}}$ is the effective capacitance coefficient of user $k$. The sensed data handled by user $k$ locally can be expressed as ${{{f}_{k}}\alpha T}{C_{}}^{-1}$, where $C$ denotes the average cycles required for computing 1-bit of data. Based on the analysis above, the total completed sensed data handled by user $k$ during the offloading process can be expressed as
\begin{equation}
  \label{eq12}
  R_{k}^{\text{p}}=r_{k}^{\text{p}}\alpha T+{{{f}_{k}}\alpha T}{C_{}}^{-1}.
\end{equation}

Based on the “\emph{sense-then-offload}” mode, all devices should sense data before computation offloading, which means that the upper limit on the amount of completed data for all devices during the second phase is limited by the amount of sensed data during the first phase. This introduces the following sensing data constraints regarding user $k$ and RIS $k$.
\begin{align}
  \label{eq12-1}
  R_{k}^{\text{p}}\le M_{k}^{\text{p}},\\
  \label{eq12-2}
  R_{k}^{\text{s}}\le M_{k}^{\text{s}}.
\end{align}

\subsection{Problem Formulation}
In this paper, our aim is to maximize the total completed sensed bits of all users and RISs by jointly optimizing the time allocation parameter, the passive beamforming at each RIS, the transmit beamforming at BS, and the energy partition parameters for all users under the constraints on the size of sensed data $\{ M_{k}^{\text{p}}, M_{k}^{\text{s}}\} $, energy supply $E_{k}^{\text{max}}$ and given time cycle $T$. Define the matrices $\mathbf{\Theta}  =\left[ {\mathbf{\Theta}_{1}},{\mathbf{\Theta}_{2}},\cdots ,{\mathbf{\Theta}_{K}} \right]$, $\mathbf{W}=\left[ {\mathbf{w}_{1}},{\mathbf{w}_{2}},\cdots ,{\mathbf{w}_{K}} \right]$, and the vector $\boldsymbol{\beta } =\left[ {{\beta }_{1}},{{\beta }_{2}},\cdots ,{{\beta }_{K}} \right]$, the corresponding total completed sensed task maximization problem can be mathematically formulated as
\begin{equation}
  \label{p1}
  \begin{aligned}
    &\underset{\mathbf{\Theta} ,\mathbf{W} ,\boldsymbol{\beta} ,\alpha }{\mathop{\max }}\,f\left( \mathbf{\Theta} ,\mathbf{W} ,\boldsymbol{\beta} ,\alpha \right)=\sum\limits_{k=1}^{K}{R_{k}^{\text{p}}+R_{k}^{\text{s}}} \\
     \text{s}\text{.t}\text{.} \ & \text{C1}:{{\beta }_{k}}\in \left[ 0,1 \right],\forall k\in \mathcal{K},  \\
    & \text{C2}:\left| {{e}^{\text{j}\theta _{n}^{k}}} \right|=1,\forall k\in \mathcal{K},\forall n\in \mathcal{N}, \\ 
    & \text{C3}:\left\| {\mathbf{w}_{k}} \right\|_{2}^{2}\le 1,\forall k\in \mathcal{K}, \\ 
    & \text{C4}:(\ref{eq12-1}),(\ref{eq12-2}),  \\
  \end{aligned}
\end{equation}
where $\alpha $ is a manually adjusted variable. Constraints $\text{C1}$ and $\text{C3}$ denote the range of energy partition fraction and the transmit beamforming at BS. Constraint $\text{C2}$ denotes the unit modulus of each RIS phase shift. Constraint $\text{C4}$ denotes that the upper limit on the amount of completed data for all devices during the second phase is limited by the amount of sensed data during the first phase, and the approximate optimal time allocation parameter is obtained by considering the constraint $\text{C4}$ in the experiment section.

The objective function of Problem (\ref{p1}) contains not only the multiple coupled variables, but also the rate fraction terms that involve mathematical expectation. Additionally, the unit modulus of phase shift constraint also complicates this problem. Therefore, it is difficult to find the global optimal solution of this problem within a prohibited complexity. To address this nonconvex problem, we seek to find approximate solutions with an efficient algorithm that can solve this problem iteratively with guaranteed convergence.

\section{Problem Solutions}
In this section, we propose an AO-based algorithm to tackle Problem (\ref{p1}). Specifically, we first use Monte Carlo and Lagrangian dual transformation techniques \cite{23} to convert the original problem with a given factor $\alpha $. Then, we use FP \cite{24} and AO methods to decompose the converted problem into three subproblems and solve them by iteratively optimizing the variables.

\subsection{Problem Transformation}
The goal of Problem (\ref{p1}) is to maximize the total completed sensed bits, which is related to the size of data sensed by users and RISs in the first phase, and the amount of data tackled by users and RISs in the second phase. In addition, $\alpha$ is a weight variable that regulates the proportion of sensing time and computation offloading time. With given $\alpha$, the size of data sensed by users and RISs in the first phase and the remaining energy budget for each user can be determined, but the exact maximum amount of data addressed by users and RISs in the second phase is difficult to be computed, which is related to the optimization of other variables. Thus, we focus on exploring the maximum amount of data addressed by users and RISs in the second phase with different given $\alpha$, and the new problem is given by
\begin{equation}
  \label{p2}
  \begin{aligned}
    & \underset{\boldsymbol{\Theta} ,\mathbf{W} ,\boldsymbol{\beta} }{\mathop{\max }}\,\ f\left( \mathbf{W},\mathbf{\Theta} ,\boldsymbol{\beta}  \right)\\ 
    & \text{s}\text{.t}\text{.}\ \text{C1-C3}. \\ 
  \end{aligned}
\end{equation}
Herein, to maximize the remaining energy efficiency \cite{8,21-1}, we temporarily ignore the constraint $\text{C4}$. In addition, the effects of constraint $\text{C4}$ on system performance are explored in the experiment section. Next, to solve Problem (\ref{p2}), we use Monte Carlo and Lagrangian dual transformation techniques to approximate expectations and decouple multiple tightly coupled variables, respectively. Thus, Problem (\ref{p2}) can be equivalently rewritten as
\begin{equation}
  \label{p3}
  \begin{aligned}
    & \underset{\boldsymbol{\Theta} ,\mathbf{W} ,\boldsymbol{\beta} ,\boldsymbol{\eta} }{\mathop{\max }}\,\ f_1^{}\left( \mathbf{W},\mathbf{\Theta} ,\boldsymbol{\beta} ,\boldsymbol{\eta}  \right) \\ 
    &\text{s}\text{.t}\text{.}\ \text{C1-C3}, \\ 
  \end{aligned}
\end{equation}
where the new function $f_1^{}\left( { \mathbf{W},\mathbf{\Theta} ,\boldsymbol{\beta} ,\boldsymbol{\eta} } \right)$ can be given by
\begin{equation}
  \label{eq13}
  \begin{aligned}
    &f_1^{}\left( {{\mathbf{W},\mathbf{\Theta} ,\boldsymbol{\beta} ,\boldsymbol{\eta}} } \right)\\
    &= \sum\limits_{k = 1}^K {\sum\limits_{j = 1}^2 {\frac{{\alpha TB}}{2}{{\log }_2}\left( {1 + \eta _{kj}^{\rm{p}}} \right)} }  - \sum\limits_{k = 1}^K {\sum\limits_{j = 1}^2 {\frac{{\alpha TB}}{2}\eta _{kj}^{\rm{p}}} } \\
    &+ \sum\limits_{k = 1}^K {\frac{{\alpha TB}}{Q}{{\log }_2}\left( {1 + \eta _k^{\rm{s}}} \right)}  - \sum\limits_{k = 1}^K {\frac{{\alpha TB}}{Q}\eta _k^{\rm{s}}} \\
    &+ \sum\limits_{k = 1}^K {\sum\limits_{j = 1}^2 {\frac{{\alpha TB}}{2}\left( {1 + \eta _{kj}^{\rm{p}}} \right)} } \frac{{{{\left| {\sqrt {{p_k}} \mathbf{w}_k^{\rm{H}}\left( {{\mathbf{h}_{{\rm{d}},k}} + {{\tilde c}_j}{\mathbf{b}_k}} \right)} \right|}^2}}}{{{{\left| {\sqrt {{p_k}} \mathbf{w}_k^{\rm{H}}\left( {{\mathbf{h}_{{\rm{d}},k}} + {{\tilde c}_j}{\mathbf{b}_k}} \right)} \right|}^2} + {n_k}}}\\
    &+ \sum\limits_{k = 1}^K {\alpha TB\left( {1 + \eta _k^{\rm{s}}} \right)\frac{{{{\left| {\sqrt {{p_k}} \mathbf{w}_k^{\rm{H}}{\mathbf{b}_k}} \right|}^2}}}{{Q{{\left| {\sqrt {{p_k}} \mathbf{w}_k^{\rm{H}}{\mathbf{b}_k}} \right|}^2} + {n_k}}}} \\
    &+ \sum\limits_{k = 1}^K {\frac{{\alpha T}}{C}\sqrt[3]{{\frac{{(1 - {\beta _k})E_k^{\rm{o}}}}{{\alpha T{\kappa _l}}}}}},
    \end{aligned}     
\end{equation}

In Eq. (\ref{eq13}), $\boldsymbol{\eta }=\left\{ \eta _{kj}^{\text{p}},\eta _{k}^{\text{s}},k\in \mathcal{K},j\in \left\{ 1,2 \right\} \right\}$ is a collection of auxiliary variables for decoding the SINR of Eq. (\ref{eq5}) and Eq. (\ref{eq7}), ${{\tilde{c}}_{1}}=-1$, ${\tilde c_2}=1$, and ${n_k} = \sum\nolimits_{i = k + 1}^K {{{\left| {\sqrt {{p_i}} \mathbf{w}_k^{\rm{H}}{\mathbf{h}_{{\rm{d}},i}}} \right|}^2}}  + \sum\nolimits_{i = k + 1}^K {{{\left| {\sqrt {{p_i}} \mathbf{w}_k^{\rm{H}}{\mathbf{b}_i}} \right|}^2}}  + \sigma _k^2$.

In Problem (\ref{p3}), when $\mathbf{W}$, $\mathbf{\Theta}$ and $\boldsymbol{\beta}$ hold fixed, it is easy to show that $f_1^{}\left( { \mathbf{W},\mathbf{\Theta} ,\boldsymbol{\beta} ,\boldsymbol{\eta} } \right)$ is a convex function with respect to $\boldsymbol{\eta }$. By setting $\partial f_{1}^{{}}\left( \mathbf{W},\mathbf{\Theta} ,\boldsymbol{\beta} , \boldsymbol{\eta}  \right)/\partial \eta _{kj}^{\text{p}}$ and $\partial f_{1}^{{}}\left( \mathbf{W},\mathbf{\Theta} ,\boldsymbol{\beta} , \boldsymbol{\eta}  \right)/\partial \eta _{k}^{\text{s}}$ to zero, the optimal $\boldsymbol{\eta }^{\text{op}}$ can be given by
\begin{equation}
  \label{eq14}
  {\boldsymbol{\eta }^{\text{op}}}=\left\{ \begin{matrix}
    {{\left( \eta _{kj}^{\text{p}} \right)}^{\text{op}}}=\frac{{{\left| \sqrt{{{p}_{k}}}\mathbf{w}_{k}^{\text{H}}\left( {\mathbf {h}_{\text{d},k}}+{{{\tilde{c}}}_{j}}{\mathbf{b}_{k}} \right) \right|}^{2}}}{{{n}_{k}}},  \\
    {{\left( \eta _{k}^{\text{s}} \right)}^{\text{op}}}=\frac{Q{{\left| \sqrt{{{p}_{k}}}\mathbf{w}_{k}^{\text{H}}{\mathbf{b}_{k}} \right|}^{2}}}{{{n}_{k}}}.  \\
 \end{matrix} \right\}_{k\in \mathcal{K} \hfill\atop j\in \left\{ 1,2 \right\}}^{} 
\end{equation}

With the optimal ${\boldsymbol{\eta }^{\text{op}}}$, the objective function of Problem (\ref{p3}) can be converted into a solvable one. Next, we will introduce the specific details of using the AO algorithm to optimize the $\mathbf{W}$, $\mathbf{\Theta} $, and $\boldsymbol{\beta} $ alternatively.

\subsection{Receive Beamforming Design}
Here, we optimize the $\mathbf{W}$ with given $\boldsymbol{\eta }$, $\mathbf{\Theta}$, and $\boldsymbol{\beta}$. As shown in Eq. (\ref{eq13}), the 5th and 6th nonconvex fractional terms in $f_1^{}\left( { \mathbf{W},\mathbf{\Theta} ,\boldsymbol{\beta} ,\boldsymbol{\eta} } \right)$ are associated with $\mathbf{W}$, which complicates the subproblem of optimizing $\mathbf{W}$. Thus, we resort to FP technique (quadratic transformation) \cite{24} to tackle the fractional terms, then the subproblem of optimizing $\mathbf{W}$ can be expressed as
\begin{equation}
  \label{p4}
  \begin{aligned}
    & \underset{\mathbf{W},\boldsymbol{\gamma} }{\mathop{\max }}\,\;{{f}_{2}}\left( \mathbf{W},\boldsymbol{\gamma}  \right) \\ 
   & \text{s}\text{.t}\text{.}\;\text{C3}, \\ 
  \end{aligned}  
\end{equation}
where $\boldsymbol{\gamma }=\left\{ \gamma _{kj}^{\text{p}},\gamma _{k}^{\text{s}},k\in \mathcal{K},j\in \left\{ 1,2 \right\} \right\}$ is a collection of auxiliary variables for transforming nonconvex terms, and ${{f}_{2}}\left( \mathbf{W},\boldsymbol{\gamma}  \right)$ is given by
\begin{equation}
  \label{eq15}
  \begin{aligned}
    &f_2^{}\left( {\mathbf{W},\boldsymbol{\gamma} } \right)\\
    &=\sum\limits_{k = 1}^K {\left[ { - {{\left| {\gamma _k^{\rm{s}}} \right|}^2}\left( {Q{{\left| {\sqrt {{p_k}} \mathbf{w}_k^{\rm{H}}{\mathbf{b}_k}} \right|}^2} + {n_k}} \right)} \right.} \\
    &+\left. {2{\mathop{\rm Re}\nolimits} \left\{ {{{\left( {\gamma _k^{\rm{s}}} \right)}^*}\sqrt {{p_k}\alpha TB\left( {1 + \eta _k^{\rm{s}}} \right)} \mathbf{w}_k^{\rm{H}}{\mathbf{b}_k}} \right\}} \right]\\
    &+\sum\limits_{k = 1}^K {\sum\limits_{j = 1}^2 {\left[ -{{\left| {\gamma _{kj}^{\rm{p}}} \right|}^2}\left( {{{\left| {\sqrt {{p_k}} \mathbf{w}_k^{\rm{H}}\left( {{\mathbf{h}_{{\rm{d}},k}} + {{\tilde c}_j}{\mathbf{b}_k}} \right)} \right|}^2} + {n_k}} \right) \right.} } \\
    &+\left. {2{\mathop{\rm Re}\nolimits} \left\{ {{{\left( {\gamma _{kj}^{\rm{p}}} \right)}^*}\sqrt {\frac{{{p_k}\alpha TB\left( {1 + \eta _{kj}^{\rm{p}}} \right)}}{2}} \mathbf{w}_k^{\rm{H}}\left( {{\mathbf{h}_{{\rm{d}},k}} + {{\tilde c}_j}{\mathbf{b}_k}} \right)} \right\}} \right]\\
    &+{\rm{const1}},
  \end{aligned}    
\end{equation}

In Eq. (\ref{eq15}), ${\rm{const1}}$ is a constant that is irrelevant to $\mathbf{W}$ and $\boldsymbol{\gamma}$. Similar to Eq. (\ref{eq14}), when $\mathbf{W}$ holds fixed, it is also easy to prove that ${{f}_{2}}\left( \mathbf{W},\boldsymbol{\gamma}  \right)$ is convex with respect to $\boldsymbol{\gamma}$. By setting $\partial f_{2}^{{}}\left( \mathbf{W},\boldsymbol{\gamma}  \right)/\partial \gamma _{kj}^{\text{p}}$ and $\partial f_{2}^{{}}\left( \mathbf{W},\boldsymbol{\gamma}  \right)/\partial \gamma _{k}^{\text{s}}$ to zero, the optimal ${{\boldsymbol{\gamma }}^{\text{op}}}$ can be obtained as
\begin{equation}
  \label{eq16}
  {\boldsymbol{\gamma} ^{{\rm{op}}}} = \left\{ {\begin{array}{*{20}{c}}
    {{{\left( {\gamma _{kj}^{\rm{p}}} \right)}^{{\rm{op}}}} = \frac{{\sqrt {\frac{{{p_k}\alpha TB\left( {1 + \eta _{kj}^{\rm{p}}} \right)}}{2}} \mathbf{w}_k^{\rm{H}}\left( {{\mathbf{h}_{{\rm{d}},k}} + {{\tilde c}_j}{\mathbf{b}_k}} \right)}}{{{{\left| {\sqrt {{p_k}} \mathbf{w}_k^{\rm{H}}\left( {{\mathbf{h}_{{\rm{d}},k}} + {{\tilde c}_j}{\mathbf{b}_k}} \right)} \right|}^2} + {n_k}}}},\\
    {{{\left( {\gamma _k^{\rm{s}}} \right)}^{{\rm{op}}}} = \frac{{\sqrt {{p_k}\alpha TB\left( {1 + \eta _k^{\rm{s}}} \right)} \mathbf{w}_k^{\rm{H}}{\mathbf{b}_k}}}{{Q{{\left| {\sqrt {{p_k}} \mathbf{w}_k^{\rm{H}}{\mathbf{b}_k}} \right|}^2} + {n_k}}}}.
  \end{array}} \right\}_{k\in \mathcal{K} \hfill\atop j\in \left\{ 1,2 \right\}}^{}    
\end{equation}

After obtaining the optimal $\boldsymbol{\gamma }^{\text{op}}$, ${{f}_{2}}\left( \mathbf{W},\boldsymbol{\gamma}  \right)$ is also convex with respect to $\mathbf{W}$, and the optimal $\mathbf{w}_k^{{\rm{op}}}$ can be obtained by
\begin{equation}
  \label{eq17}
  \mathbf{w}_k^{{\rm{op}}} = \left\{ {\begin{array}{*{20}{c}}
    {{\bf{B}}_k^{ - 1}{{\bf{a}}_k},}&{\left\| {{\bf{B}}_k^{ - 1}{{\bf{a}}_k}} \right\| \le 1},\\
    {{\bf{B}}_k^{ - 1}{{\bf{a}}_k}/\left\| {{\bf{B}}_k^{ - 1}{{\bf{a}}_k}} \right\|,}&{{\rm{otherwise}}},
    \end{array}} \right.    
\end{equation}
where ${{\mathbf{a}}_{k}} = \sum\nolimits_{j = 1}^2 {{{\left( {\gamma _{kj}^{\rm{p}}} \right)}^*}\sqrt {\frac{{{p_k}\alpha TB\left( {1 + \eta _{kj}^{\rm{p}}} \right)}}{2}} \left( {{{\mathbf{h}}_{{\rm{d}},{k}}} + {{\tilde c}_j}{{\mathbf{b}}_{k}}} \right) + }$ ${\left( {\gamma _k^{\rm{s}}} \right)^*}\sqrt {{p_k}\alpha TB\left( {1 + \eta _k^{\rm{s}}} \right)} {{\bf{b}}_{k}}$, ${{{\mathbf{B}}_{k}} = \sum\nolimits_{j = 1}^2 {{{\left| {\gamma _{kj}^{\rm{p}}} \right|}^2}\left[ {{p_k}{\mathbf{h}_k}\mathbf{h}_k^{\rm{H}}} \right.}  + }$ $\left. { + {\boldsymbol{\delta}_{k}}} \right] + {\left| {\gamma _k^{\rm{s}}} \right|^2}\left[ {Q{p_k}{{\mathbf{b}}_{k}}\mathbf{b}_k^{\rm{H}} + {\boldsymbol{\delta} _{k}}} \right]$, where ${\mathbf{h}_{k}}={\mathbf{h}_{\text{d},k}}+{{\tilde{c}}_{j}}{\mathbf{b}_{k}}$, ${\boldsymbol{\delta } _k} = \sum\nolimits_{i = k + 1}^K {{p_i}\left( {{\mathbf{h}_{{\rm{d}},i}}\mathbf{h}_{{\rm{d}},i}^{\rm{H}} + {\mathbf{b}_i}\mathbf{b}_i^{\rm{H}}} \right)}  + {\sigma ^2}{\mathbf{I}_M}$, and $\mathbf{I}_M $ is the diagonal matrix of 1.

In this subsection, we can obtain the optimal ${\boldsymbol{\gamma } ^{{\rm{op}}}}$ and $\mathbf{w}_{k}^{\text{op}}$ according to Eq. (\ref{eq16}) and Eq. (\ref{eq17}), respectively. In the next subsection, $\mathbf{w}_{k}^{\text{op}}$ will be used for the $\mathbf{\Theta}$ optimization problem.

\subsection{RIS Reflecting Shift Design}
In this subsection, an iterative method is utilized to optimize the RIS reflecting shifts with given $\mathbf{W}$, $\boldsymbol{\beta}$ and $\boldsymbol{\eta }$. The basic idea of this method is that the rest reflecting shifts are fixed when one of reflecting shift is optimizing. Following a similar procedure to optimize $\mathbf{W}$, we first use FP technique to transform the Problem (\ref{p3}). As a result, the subproblem of optimizing $\mathbf{\Theta}$ can be equivalently transformed to
\begin{equation}
  \label{p5}
  \begin{aligned}
   & \underset{\boldsymbol{\Theta} ,\boldsymbol{\xi} }{\mathop{\max }}\,\;f_{3}^{{}}\left( \boldsymbol{\Theta} ,\boldsymbol{\xi}  \right) \\ 
   & \text{s}\text{.t}\text{.}\;\text{C2}, \\ 
  \end{aligned}  
\end{equation}
where $\boldsymbol{\xi }=\left\{ \xi _{kj}^{\text{p}},\xi _{k}^{\text{s}},k\in \mathcal{K},j\in \left\{ 1,2 \right\} \right\}$ is a collection of auxiliary variables for transforming nonconvex terms, and ${{f}_{3}}\left( \boldsymbol{\Theta} ,\boldsymbol{\xi} \right)$ is given by
\begin{equation}
  \label{eq18}
  \begin{aligned}
    &f_3^{}\left( {\boldsymbol{\Theta} ,\boldsymbol{\xi} } \right)\\
    &= \sum\limits_{k = 1}^K {\left[ { - {{\left| {\xi _k^{\rm{s}}} \right|}^2}\left( {Q{{\left| {\mathbf{h}_{kk}^{{\rm{RIS}}}{\boldsymbol{\varphi} _k}} \right|}^2} + {n_k}} \right)} \right.} \\
    &+\left. {2{\mathop{\rm Re}\nolimits} \left\{ {{{\left( {\xi _k^{\rm{s}}} \right)}^*}\sqrt {\alpha TB\left( {1 + \eta _k^{\rm{s}}} \right)} \mathbf{h}_{kk}^{{\rm{RIS}}}{\boldsymbol{\varphi} _k}} \right\}} \right]\\
    &+\sum\limits_{k = 1}^K {\sum\limits_{j = 1}^2 {\left[ { - {{\left| {\xi _{kj}^{\rm{p}}} \right|}^2}\left( {{{\left| {\left( {{h}_{kk}^{\rm{d}} + {{\tilde c}_j}\mathbf{h}_{kk}^{{\rm{RIS}}}{\boldsymbol{\varphi} _k}} \right)} \right|}^2} + {n_k}} \right)} \right.} } \\
    &+\left. {2{\mathop{\rm Re}\nolimits} \left\{ {{{\left( {\xi _{kj}^{\rm{p}}} \right)}^*}\sqrt {\frac{{\alpha TB\left( {1 + \eta _{kj}^{\rm{s}}} \right)}}{2}} \left( {{h}_{kk}^{\rm{d}} + {{\tilde c}_j}\mathbf{h}_{kk}^{{\rm{RIS}}}{\boldsymbol{\varphi} _k}} \right)} \right\}} \right]\\
    &+{\rm{const2}},
    \end{aligned}
\end{equation}

In Eq. (\ref{eq18}), $\mathbf{h}_{ki}^{{\rm{RIS}}} = \sqrt {{p_k}} { {\mathbf{w}_k^{\rm{H}}\mathbf{G}_i^{\rm{H}}{\rm{diag}}\left( {{\mathbf{h}_{{\rm{r}},i}}} \right)} }$, ${h}_{ki}^{\rm{d}} = \sqrt {{p_k}} \mathbf{w}_k^{\rm{H}}{\mathbf{h}_{{\rm{d}},i}}$, ${\boldsymbol{\varphi} _k} = {\left[ {\varphi _1^k,\varphi _2^k, \cdots ,\varphi _N^k} \right]^{\rm{H}}} \in {\mathbb{C}^{N \times 1}}$, and ${\rm{const2}}$ is a constant that is irrelevant to $\boldsymbol{\Theta}$ and $\boldsymbol{\xi}$. Similar to (\ref{eq14}), when $\boldsymbol{\Theta}$ holds fixed, it is also easy to prove that ${{f}_{3}}\left( \boldsymbol{\Theta} ,\boldsymbol{\xi} \right)$ is convex with respect to $\boldsymbol{\xi}$. By setting $\partial f_{3}^{{}}\left( \boldsymbol{\Theta} ,\boldsymbol{\xi}  \right)/\partial \xi _{kj}^{\text{p}}$ and $\partial f_{3}^{{}}\left( \boldsymbol{\Theta} ,\boldsymbol{\xi}  \right)/\partial \xi _{k}^{\text{s}}$ to zero, the optimal ${{\boldsymbol{\xi }}^{\text{op}}}$ can be obtained as
\begin{equation}
  \label{eq19}
  {\boldsymbol{\xi} ^{{\rm{op}}}} = \left\{ {\begin{array}{*{20}{c}}
    {{{\left( {\xi _{kj}^{\rm{p}}} \right)}^{{\rm{op}}}} = \frac{{\sqrt {\frac{{\alpha TB\left( {1 + \eta _{kj}^{\rm{p}}} \right)}}{2}} \left( {{h}_{kk}^{\rm{d}} + {{\tilde c}_j}\mathbf{h}_{kk}^{{\rm{RIS}}}{\boldsymbol{\varphi}_k}} \right)}}{{{{\left| {\left( {{h}_{kk}^{\rm{d}} + {{\tilde c}_j}\mathbf{h}_{kk}^{{\rm{RIS}}}{\boldsymbol{\varphi} _k}} \right)} \right|}^2} + {n_k}}}},\\
    {{{\left( {\xi _k^{\rm{s}}} \right)}^{{\rm{op}}}} = \frac{{\sqrt {\alpha TB\left( {1 + \eta _k^{\rm{s}}} \right)} \mathbf{h}_{kk}^{{\rm{RIS}}}{\boldsymbol{\varphi} _k}}}{{Q{{\left| {\mathbf{h}_{kk}^{{\rm{RIS}}}{\boldsymbol{\varphi} _k}} \right|}^2} + {n_k}}}}.
  \end{array}} \right\}_{k\in \mathcal{K} \hfill\atop j\in \left\{ 1,2 \right\}}^{}    
\end{equation}

Then, with the optimal ${{\boldsymbol{\xi }}^{\text{op}}}$, ${{f}_{3}}\left( \boldsymbol{\Theta} ,\boldsymbol{\xi} \right)$ can be transformed into a simpler expression as follows
\begin{equation}
  \label{eq20}
  f_3^{}\left( \boldsymbol{\Theta}  \right) =  - \sum\limits_{k = 1}^K {\boldsymbol{\varphi} _k^{\rm{H}}{\mathbf{U}_k}{\boldsymbol{\varphi} _k}}  + \sum\limits_{k = 1}^K {2{\mathop{\rm Re}\nolimits} \left\{ {\mathbf{z}_k^{\rm{H}}{\boldsymbol{\varphi} _k}} \right\}}  + {\rm{const3}},
\end{equation}
where $\rm{const3}$ is a constant that is irrelevant to $\boldsymbol{\Theta}$,
\begin{align}
  \label{eq21}
    {\mathbf{U}_k} =& \sum\limits_{j = 1}^2 {{{\left| {\xi _{kj}^{\rm{p}}} \right|}^2}{{\left( {\mathbf{h}_{kk}^{{\rm{RIS}}}} \right)}^{\rm{H}}}\mathbf{h}_{kk}^{{\rm{RIS}}}}  + {\left| {\xi _k^{\rm{s}}} \right|^2}Q{\left( {\mathbf{h}_{kk}^{{\rm{RIS}}}} \right)^{\rm{H}}}\mathbf{h}_{kk}^{{\rm{RIS}}}\nonumber\\
                    &+ \sum\limits_{i = 1}^{k - 1} {{{\left( {\mathbf{h}_{ik}^{{\rm{RIS}}}} \right)}^{\rm{H}}}\mathbf{h}_{ik}^{{\rm{RIS}}}\left( {{{\left| {\xi _i^{\rm{s}}} \right|}^2} + \sum\limits_{j = 1}^2 {{{\left| {\xi _{ij}^{\rm{p}}} \right|}^2}} } \right)}, \\
  \label{eq22}
    {\mathbf{z}_k} =& \sum\limits_{j = 1}^2 {{{\tilde c}_j}\left( {\xi _{kj}^{\rm{p}}\sqrt {\alpha TB\left( {1 + \eta _{kj}^{\rm{s}}} \right)/2} {{\left( {\mathbf{h}_{kk}^{{\rm{RIS}}}} \right)}^{\rm{H}}} - } \right.} \nonumber\\
                    &\left. {{{\left| {\xi _{kj}^{\rm{p}}} \right|}^2}{h}_{kk}^{\rm{d}}{{\left( {\mathbf{h}_{kk}^{{\rm{RIS}}}} \right)}^{\rm{H}}}} \right) + \xi _k^{\rm{s}}\sqrt {\alpha TB\left( {1 + \eta _k^{\rm{s}}} \right)} {\left( {\mathbf{h}_{kk}^{{\rm{RIS}}}} \right)^{\rm{H}}}.
\end{align}

Next, we transform the vector items of ${{f}_{3}}\left( \boldsymbol{\Theta} \right)$ into scalar forms as in \cite{25}, which can be equivalently represented as follows
\begin{align}
  \label{eq23}
    \sum\limits_{k = 1}^K {\boldsymbol{\varphi}_k^{\rm{H}}{\mathbf{U}_k}{\boldsymbol{\varphi} _k}}  =& \varphi _{kn}^ * {u_{k,nn}}\varphi _{kn}^{} + 2{\mathop{\rm Re}\nolimits} \left\{ {\sum\limits_{i \ne n}^N {\varphi _{kn}^ * {u_{k,ni}}\varphi _{ki}^{}} } \right\} \nonumber \\
        &+ \sum\limits_{i \ne n}^N {\sum\limits_{j \ne n}^N {\varphi _{ki}^ * {u_{k,ij}}\varphi _{kj}^{}} }  \nonumber \\&+ \sum\limits_{l \ne k}^K {\sum\limits_{i = 1}^N {\sum\limits_{j = 1}^N {\varphi _{li}^ * {u_{l,ij}}\varphi _{lj}^{}} } },\\ 
  \label{eq24}
      \sum\limits_{k = 1}^K {\boldsymbol{\varphi} _k^{\rm{H}}\mathbf{z}_k^{}}  =& \varphi _{kn}^ * {z_{kn}} + \sum\limits_{i \ne n}^N {\varphi _{ki}^ * {z_{ki}}}  + \sum\limits_{j \ne k}^N {\sum\limits_{i = 1}^N {\varphi _{ji}^ * {z_{ji}}} },
\end{align}
where $\varphi _{kn}^{}$ and ${z_{kn}}$ are the $n$-element of ${{\boldsymbol{\varphi} _k}}$ and ${\mathbf{z}_k}$, respectively, and ${{u_{k,ij}}}$ is the element at $i$-th row and $j$-th of ${{\mathbf{U}_k}}$. By substituting Eq. (\ref{eq23}) and Eq. (\ref{eq24}) into ${{f}_{3}}\left( \boldsymbol{\Theta} \right)$ and removing the terms that are irrelevant to ${\varphi _{kn} }$, the function with respect to ${\varphi _{kn}}$ can be expressed as
\begin{equation}
  \label{eq25}
  f_{3{\varphi _{kn}}}^{}\left( {\varphi _{kn}^{}} \right) = 2{\mathop{\rm Re}\nolimits} \left\{ {\varphi _{kn}^ * {{{A}}_{1,kn}}} \right\} - {\left| {{\varphi _{kn}}} \right|^2}{{{A}}_{2,kn}},
\end{equation}
where ${{{A}}_{1,kn}} = {z_{kn}} - \sum\nolimits_{i \ne n}^N {{u_{k,ni}}{\varphi _{ki}}}$, and ${{{A}}_{2,kn}} = {u_{k,nn}}$. 

Thus, the optimization $\boldsymbol{\Theta}$ can be further decomposed to $KN$ subproblems, and the subproblem with respect to ${\varphi _{kn} }$ is given by
\begin{equation}
  \label{p6}
  \begin{array}{l}
    \mathop {\max }\limits_{\varphi _{kn}^{}} \;\,f_{3\varphi _{kn}^{}}^{}\left( {\varphi _{kn}^{}} \right)\\
    \begin{array}{*{20}{l}}
    {{\rm{s}}{\rm{.t}}{\rm{.}}}&{{\rm{C2}}}.
    \end{array}
    \end{array}    
\end{equation}

In addition to give the optimal solution for the continuous phase shift in the original problem, we also give the optimal solution for the discrete phase shift. The optimal solutions can be given as follows
\begin{itemize}
  \item Continuous phase shift: As for this case, ${\left| {{e^{{\rm{j}}\theta _n^k}}} \right|^2} = 1$. The optimal phase shift can be computed as
  \begin{equation}
    \label{eq26}
    {{\left( \angle \theta _{n}^{k} \right)}^{\text{opt}}}=\underset{x\in \left[ 0,2\pi  \right]}{\mathop{\arg \min }}\,\left| x-\angle {{{A}}_{1,kq}} \right|.
  \end{equation}
  \item Discrete  phase shift: As for this case, ${\left| {{e^{{\rm{j}}\theta _n^k}}} \right|^2} = 1$. The optimal phase shift can be computed as
  \begin{equation}
    \label{eq27}
    {\left( {\angle \theta _n^k} \right)^{{\rm{opt}}}} = \mathop {\arg \min }\limits_{x \in \left\{ {0,\frac{{2\pi }}{\tau }, \cdots ,\frac{{2\pi (\tau  - 1)}}{\tau }} \right\}} \left| {x - \angle {{{A}}_{1,kq}}} \right|,
  \end{equation}
  where $\tau ={{2}^{b}}$ denotes the $b$-bit discrete phase shift.
\end{itemize}

\subsection{Energy Partition Optimization}
Here, similar to subsection B and subsection C, we first use FP technique to transform the Problem (\ref{p3}) with given $\mathbf{W}$, $\boldsymbol{\Theta}$ and $\boldsymbol{\eta }$. As a result, the subproblem of optimizing $\boldsymbol{\beta }$ can be equivalently given by
\begin{equation}
  \label{p7}
  \begin{aligned}
   & \underset{\boldsymbol{\beta} ,\boldsymbol{\lambda} }{\mathop{\max }}\,\;f_{4}^{{}}\left( \boldsymbol{\beta} ,\boldsymbol{\lambda}  \right) \\ 
   & \text{s}\text{.t}\text{.}\;\text{C1}, \\ 
  \end{aligned}  
\end{equation}
where $\boldsymbol{\lambda }=\left\{ \lambda _{kj}^{\text{p}},\lambda _{k}^{\text{s}},k\in \mathcal{K},j\in \left\{ 1,2 \right\} \right\}$ is a collection of auxiliary variables for transforming nonconvex terms, and ${{f}_{4}}\left( \boldsymbol{\beta} ,\boldsymbol{\lambda} \right)$ is given by
\begin{equation}
  \label{eq28}
  \begin{aligned}
    &{f_4^{}\left( {\boldsymbol{\beta} ,\boldsymbol{\lambda} } \right)}\\
    &{ = \sum\limits_{k = 1}^K {\left[ { - {{\left| {\lambda _k^{\rm{s}}} \right|}^2}\left( {Q{{\left| {{\mathbf{w}}_k^{\rm{H}}{{\mathbf{b}}_k}} \right|}^2}{p_k}\left( {{\beta _k}} \right) + {n_k}} \right)} \right.} }\\
    &{\left. { + 2{\rm{Re}}\left\{ {{{\left( {\lambda _k^{\rm{s}}} \right)}^*}\sqrt {\alpha TB\left( {1 + \eta _k^{\rm{s}}} \right)} {\mathbf{w}}_k^{\rm{H}}{{\mathbf{b}}_k}} \right\}\sqrt {{p_k}\left( {{\beta _k}} \right)} } \right]}\\
    &{ + \sum\limits_{k = 1}^K {\sum\limits_{j = 1}^2 {\left[ { - {{\left| {\lambda _{kj}^{\rm{p}}} \right|}^2}\left( {{{\left| {{\mathbf{w}}_k^{\rm{H}}\left( {{{\mathbf{h}}_{{\rm{d}},k}} + {{\tilde c}_j}{{\mathbf{b}}_k}} \right)} \right|}^2}{p_k}\left( {{\beta _k}} \right) + {n_k}} \right)} \right.} } }\\
    &+ 2{\rm{Re}}\{ {{{( {\lambda _{kj}^{\rm{p}}} )}^*}\sqrt {\frac{{\alpha TB\left( {1 + \eta _{kj}^{\rm{s}}} \right){p_k}\left( {{\beta _k}} \right)}}{2}} {\mathbf{w}}_k^{\rm{H}}\left( {{{\mathbf{h}}_{{\rm{d}},k}}} \right.} \\
    &+ \left. {\left. {\left. {{{\tilde c}_j}{{\mathbf{b}}_k}} \right)} \right\}} \right]{\rm{ + }}\;{\rm{const3}},
  \end{aligned}
\end{equation}

In Eq. (\ref{eq28}), ${\rm{const3}}$ is a constant that is irrelevant to $\boldsymbol{\beta }$ and $\boldsymbol{\lambda}$. Similar to Eq. (\ref{eq14}), when $\boldsymbol{\beta}$ holds fixed, it is also easy to prove that ${{f}_{4}}\left( \boldsymbol{\beta },\boldsymbol{\lambda} \right)$ is convex with respect to $\boldsymbol{\lambda}$. By setting $\partial f_{4}^{{}}\left( \boldsymbol{\beta} ,\boldsymbol{\lambda}  \right)/\partial \lambda _{kj}^{\text{p}}$ and $\partial f_{4}^{{}}\left( \boldsymbol{\beta} ,\boldsymbol{\lambda}  \right)/\partial \lambda _{k}^{\text{s}}$ to zero, the optimal ${{\boldsymbol{\lambda }}^{\text{op}}}$ can be obtained as
\begin{equation}
  \label{eq29}
  {\boldsymbol{\lambda} ^{{\rm{op}}}} = \left\{ {\begin{array}{*{20}{c}}
  {{{\left( {\lambda _{kj}^{\rm{p}}} \right)}^{{\rm{op}}}} = \frac{{\sqrt {\frac{{{p_k}\alpha TB\left( {1 + \eta _{kj}^{\rm{p}}} \right)}}{2}} \mathbf{w}_k^{\rm{H}}\left( {{\mathbf{h}_{{\rm{d}},k}} + {{\tilde c}_j}{\mathbf{b}_k}} \right)}}{{{{\left| {\sqrt {{p_k}} \mathbf{w}_k^{\rm{H}}\left( {{\mathbf{h}_{{\rm{d}},k}} + {{\tilde c}_j}{\mathbf{b}_k}} \right)} \right|}^2} + {n_k}}}},\\
  {{{\left( {\lambda _k^{\rm{s}}} \right)}^{{\rm{op}}}} = \frac{{\sqrt {{p_k}\alpha TB\left( {1 + \eta _k^{\rm{s}}} \right)} \mathbf{w}_k^{\rm{H}}{\mathbf{b}_k}}}{{Q{{\left| {\sqrt {{p_k}} \mathbf{w}_k^{\rm{H}}{\mathbf{b}_k}} \right|}^2} + {n_k}}}}.
  \end{array}} \right\}_{k\in \mathcal{K} \hfill\atop j\in \left\{ 1,2 \right\}}^{}
\end{equation}

Then, with the optimal ${{\boldsymbol{\lambda }}^{\text{op}}}$ and performing some mathematical transformations of ${{f}_{4}}\left( \boldsymbol{\beta} ,\boldsymbol{\lambda} \right)$, it is easy shown that the optimization of ${{\beta }_{k}}$ is independent of each other. As a result, Problem (\ref{p7}) can be decomposed into $K$ subproblems, and the $k$-subproblem for optimizing ${{\beta }_{k}}$ is given by
\begin{equation}
  \label{p8}
  \begin{array}{l}
    \mathop {\max }\limits_{{\beta _k}} \;\,f_{4{\beta _k}}^{}\left( {{\beta _k}} \right)\\
    \begin{array}{*{20}{l}}
    {{\rm{s}}{\rm{.t}}{\rm{.}}}&{{\rm{C1}}},
    \end{array}
  \end{array}    
\end{equation}
where
\begin{align}
  \label{eq30}
    f_{4{\beta _k}}^{}\left( {{\beta _k}} \right) =&  - \left( {{{{A}}_{1k}} + {{{A}}_{2k}}} \right){p_k}\left( {{\beta _k}} \right) + \left( {{{{B}}_{1k}} + {{{B}}_{2k}}} \right)\sqrt {{p_k}\left( {{\beta _k}} \right)} \nonumber\\
    &+ \frac{{\alpha T}}{C}\sqrt[3]{{\frac{{(1 - {\beta _k})E_k^{\rm{o}}}}{{\alpha T{\kappa}_{l} }}}} + {\rm{const4}},
\end{align}
where ${\rm{const4}}$ is a constant in ${{f}_{4}}\left( \boldsymbol{\beta },\boldsymbol{\lambda} \right)$ that is irrelevant to ${\beta _k}$, and
\begin{align}
  \label{eq31}
  {{{A}}_{1k}} =& {\left| {\lambda _k^{\rm{s}}} \right|^2}\left( {Q{{\left| {\mathbf{w}_k^{\rm{H}}{\mathbf{b}_k}} \right|}^2} + \sum\limits_{i = 1}^{k - 1} {{{\left| {\mathbf{w}_i^{\rm{H}}{\mathbf{h}_{{\rm{d}},k}}} \right|}^2} + {{\left| {\mathbf{w}_i^{\rm{H}}{\mathbf{b}_k}} \right|}^2}} } \right),\\
  \label{eq32}
  {{{A}}_{2k}} =& \sum\limits_{j = 1}^2 {  {{\left| {\lambda _{kj}^{\rm{p}}} \right|}^2}\left[ {{{\left| {\mathbf{w}_k^{\rm{H}}\left( {{\mathbf{h}_{{\rm{d}},k}} + {{\tilde c}_j}{\mathbf{b}_k}} \right)} \right|}^2} + } \right.}\nonumber \\
                   &\left. {\sum\limits_{i = 1}^{k - 1} {{{\left| {\mathbf{w}_i^{\rm{H}}{\mathbf{h}_{{\rm{d}},k}}} \right|}^2} + {{\left| {\mathbf{w}_i^{\rm{H}}{\mathbf{b}_k}} \right|}^2}} }\right],  \\
% \end{align}
% \begin{align}
  \label{eq33} 
  {{{B}}_{1k}} =& 2{\mathop{\rm Re}\nolimits} \left\{ {{{\left( {\lambda _k^{\rm{s}}} \right)}^*}\sqrt {\alpha TB\left( {1 + \eta _k^{\rm{s}}} \right)} \mathbf{w}_k^{\rm{H}}{\mathbf{b}_k}} \right\},\\
  \label{eq34}
  {{{B}}_{2k}} =& \sum\limits_{j = 1}^2 {2{\mathop{\rm Re}\nolimits} \left\{ {{{\left( {\lambda _{kj}^{\rm{p}}} \right)}^*}\sqrt {\alpha TB\left( {1 + \eta _{kj}^{\rm{s}}} \right)/2} \mathbf{w}_k^{\rm{H}}\left( {{\mathbf{h}_{{\rm{d}},k}}} \right.} \right.} \nonumber\\
                   &\left. {\left. { + {{\tilde c}_j}{\mathbf{b} _k}} \right)} \right\}.
\end{align}
\textit{Proposition 1:} $f_{4{\beta _k}}^{}\left( {{\beta _k}} \right)$ is concave with respect to ${\beta _k}$.

The detailed proof of Proposition 1 can be seen in Appendix.

Since $f_{4{\beta _k}}^{}\left( {{\beta _k}} \right)$ is concave with respect to ${\beta _k}$ and the constraint C2 is a convex set, Problem (\ref{p8}) can be easily solved by CVX-tools \cite{cvx}. By solving Problem (\ref{p8}), the optimal $\beta _{k}^{\text{opt}}$ can be ultimately obtained.

\subsection{Overall of Proposed Algorithm}
The proposed algorithm based on AO method for solving the total completed sensed data maximization problem in (\ref{p1}) is summarized in \textbf{Algorithm 1}. Based on the above analysis, the subproblems of optimizing $\mathbf{\Theta}$, $\mathbf{W}$, and $\boldsymbol{\beta}$ can be obtained local optimal values in each iteration. The convergence of FP technique and SDR-based phase shift optimization can be proved in \cite{24} and Eq. (\ref{eq38}), respectively, which guarantees that the objective function is monotonically nondecreasing.

Next, we discuss the complexity of Algorithm 1. The computational complexity mainly depends on the updating auxiliary variables, the optimization of RIS reflecting shifts, the transmit beamforming at BS, and  the energy partition for each user. In each iteration, $n_k$ is updated three times, the complexity of which is $O\left( \frac{3K(K-1)M}{2} \right)$. With given $n_k$, the complexity of updating $\boldsymbol{\eta}$, $\boldsymbol{\gamma}$, $\boldsymbol{\xi}$ and $\boldsymbol{\lambda}$ is $O\left( 20KM \right)$, the complexity of updating $\mathbf{W}$ is $O\left( K{{M}^{3}}+21K{{M}^{2}}+13KM \right)$, the complexity of updating $\boldsymbol{\Theta}$ based on alternating method is $O\left( 3K{{N}^{2}} \right)$, and the complexity of updating $\boldsymbol{\beta}$ is $O\left( {{K}^{4.5}} \right)$. As a result, the computational complexity of Algorithm 1 based on iteration method is $O\left( {{L}_{o}}\left( {{C}_{0}}+3K{{N}^{2}}+{{K}^{4.5}} \right) \right)$, where ${{C}_{0}}=K{{M}^{3}}+3.5{{K}^{2}}M+21K{{M}^{2}}+31.5KM$, and ${{L_o}}$ is the total iteration number of the proposed AO algorithm.

\begin{algorithm}[!t]
  \caption{The AO-based Algorithm for Solving the Original Problem (\ref{p1})}\label{alg:alg1}
  \begin{algorithmic}
  \STATE 
  \STATE {\textbf{Step 1: } Initialize $\mathbf{\Theta}$, $\mathbf{W}$, $\boldsymbol{\beta}$, $\alpha$ to feasible values.}
  \STATE {\textbf{repeat }}
  \STATE \hspace{0.5cm}{\textbf{Step 2: }Update $\boldsymbol{\eta}$ with (\ref{eq14}).}
  \STATE \hspace{0.5cm}{\textbf{Step 3.1: }Update $\boldsymbol{\gamma}$ with (\ref{eq16}).}
  \STATE \hspace{0.5cm}{\textbf{Step 3.2: }Update $\mathbf{W}$ with (\ref{eq17}).}
  \STATE \hspace{0.5cm}{\textbf{Step 4.1: }Update $\boldsymbol{\xi}$ with (\ref{eq19}).}
  \STATE \hspace{0.5cm}{\textbf{Step 4.2: }Update $\theta_{kq}$ with (\ref{eq26}) and (\ref{eq27}).}
  \STATE \hspace{0.5cm}{\textbf{Step 5.1: }Update $\boldsymbol{\lambda}$ with (\ref{eq29}).}
  \STATE \hspace{0.5cm}{\textbf{Step 5.2: }Update $\beta_k$ by solving Problem (\ref{p8}) via CVX.}
  \STATE {\textbf{until } the objective function in (\ref{p3}) converges.}
  \end{algorithmic}
  \label{alg1}
\end{algorithm}

\subsection{Benchmark with Semidefinite Relaxation Method (SDR) for RIS Reflecting Shift}
In this subsection, we propose an SDR-based method introduced in \cite{27} to optimize $\boldsymbol{\Theta } $ for comparing the iterative method in subsection C. From Eq. (\ref{eq20}), it is easy to show that the optimization of RIS reflecting shift can be decomposed into $K$ independent subproblems, and the $k$-th subproblem can be written as
\begin{equation}
  \label{p9}
  \begin{array}{l}
    \mathop {\max }\limits_{{\boldsymbol{\varphi} _k}} \;\,f_{3{\boldsymbol{\varphi} _k}}^{}\left( {{\boldsymbol{\varphi} _k}} \right) =  - \boldsymbol{\varphi} _k^{\rm{H}}{\mathbf{U}_k}{\boldsymbol{\varphi} _k} + 2{\mathop{\rm Re}\nolimits} \left\{ {\mathbf{z}_k^{\rm{H}}{\boldsymbol{\varphi} _k}} \right\}\\
    \begin{array}{*{20}{l}}
    {{\rm{s}}{\rm{.t}}{\rm{.}}}&{{\rm{C2}}},
    \end{array}
  \end{array}    
\end{equation}
where ${\mathbf{U}_{k}}$ is a positive semidefinite matrix as in Eq. (\ref{eq21}). Obviously, $f_{3{{\boldsymbol{\varphi} }_{k}}}^{{}}\left( {{\boldsymbol{\varphi} }_{k}} \right)$ is a concave function with respect to ${{\boldsymbol{\varphi} }_{k}}$ and the unit-modulus constraint $\text{C2}$ makes this subproblem nonconvex. To tackle the unit-modulus constraint, we use SDR technique to solve it. In particular, by defining a matrix ${\mathbf{Q}_{k}}\in {{\mathbb{C}}^{(N+1)\times (N+1)}}$ as
\begin{equation}
  \label{eq37}
  {\mathbf{Q}_{k}}=\left[ \begin{matrix}
    -{\mathbf{U}_{k}} & {\mathbf{z}_{k}}  \\
    \mathbf{z}_{k}^{\text{H}} & 0  \\
  \end{matrix} \right],
\end{equation}
and a vector ${\mathbf{v}_{k}}={{\left[ \varphi _{1}^{k},\varphi _{2}^{k},\ldots ,\varphi _{N}^{k},{{\rho }_{k}} \right]}^{\text{T}}}\in {{\mathbb{C}}^{(N+1)\times 1}}$ with an auxiliary scalar ${{\rho }_{k}}$. Then, $f_{3{{\boldsymbol{\varphi} }_{k}}}^{{}}\left( {{\boldsymbol{\varphi} }_{k}} \right)$ can be re-expressed as $f_{3{{\boldsymbol{\varphi} }_{k}}}^{{}}\left( {{\boldsymbol{\varphi} }_{k}} \right)=\text{tr}\left( {\mathbf{Q}_{k}}{\mathbf{V}_{k}} \right)$, where ${\mathbf{V}_{k}}={\mathbf{v}_{k}}\mathbf{v}_{k}^{\text{H}}$. Consequently, the rank-one constraint $\text{C2}$ is replaced by $\{{\mathbf{V}_{k}}\underline{\succ }0,\ \text{rank(}{\mathbf{V}_{k}}\text{) = 1,}\ \text{diag(}{\mathbf{V}_{k}}{\text{) = }}{\mathbf{I}_{N+1}}\}$. Then, by temporarily neglecting the rank-one constraint on ${\mathbf{V}_{k}}$, the $k$-th subproblem can be rewritten as
\begin{equation}
  \label{p10}
  \begin{array}{l}
    \mathop {\max }\limits_{{\mathbf{V}_k}} \;\,f_{3{\mathbf{V}_k}}^{}\left( {{\mathbf{V}_k}} \right) = {\rm{tr}}\left( {{\mathbf{Q}_k}{\mathbf{V}_k}} \right)\\
    \begin{array}{*{20}{l}}
    {{\rm{s}}{\rm{.t}}{\rm{.}}}&{{\mathbf{V}_k}\succeq 0,\;{\rm{diag(}}{\mathbf{V}_k}{\text{)  =  }}{\mathbf{I}_{N+1}}}.
    \end{array}
    \end{array}        
\end{equation}

It is noted that Problem (\ref{p9}) is relaxed to a convex semidefinite program Problem (\ref{p10}), which can be solved by CVX-tools \cite{cvx}. However, the solution ${{\mathbf{V}}_k}$ obtained by solving Problem (\ref{p10}) may not satisfy the rank-one constraint, and then we use Gaussian randomization technique \cite{28} to obtain the ${{\boldsymbol{\varphi }}_{k}}$ from ${{\mathbf{V}}_k}$. Note that the SDR technique cannot guarantee that the value of the objective function increases with the number of iterations. Therefore, to ensure the convergence of Problem (\ref{p9}), we adopt the following update
\begin{equation}
  \label{eq38}
  \boldsymbol{\varphi} _{k}^{t+1}=\left\{ \begin{matrix}
    {{{\tilde{\boldsymbol{\varphi} }}}_{k}}, & f_{3{{\boldsymbol{\varphi} }_{k}}}^{{}}\left( {{{\tilde{\boldsymbol{\varphi} }}}_{k}} \right)>f_{3{{\boldsymbol{\varphi} }_{k}}}^{{}}\left( \boldsymbol{\varphi} _{k}^{t} \right),  \\
    \boldsymbol{\varphi} _{k}^{t}, & \text{otherwise,}  \\
 \end{matrix} \right. 
\end{equation}
where $\boldsymbol{\varphi} _{k}^{t}$ denotes the value of ${{\boldsymbol{\varphi} }_{k}}$ in the $t$-th iteration, and ${{\tilde{\boldsymbol{\varphi} }}_{k}}$ denotes the suboptimal solution of Problem (\ref{p10}) in the ($t+1$)-th iteration.

Next, we analyze the complexity of the SDR-based algorithm. Compared with Algorithm 1, the difference of the benchmark is the method of updating the phase shift. Based on the SDR method, the complexity of updating $\mathbf{\Theta }$ is $O\left( 2K{{N}^{2}}+K{{N}^{3.5}} \right)$. Thus, the total complexity of the SDR-based algorithm is $O\left( {{L}_{o}}\left( {{C}_{0}}+2K{{N}^{2}}+K{{N}^{3.5}}+{{K}^{4.5}} \right) \right)$.

\section{Numerical Results}
\begin{figure}[!t]
  \centering
  \includegraphics[width=3.5in]{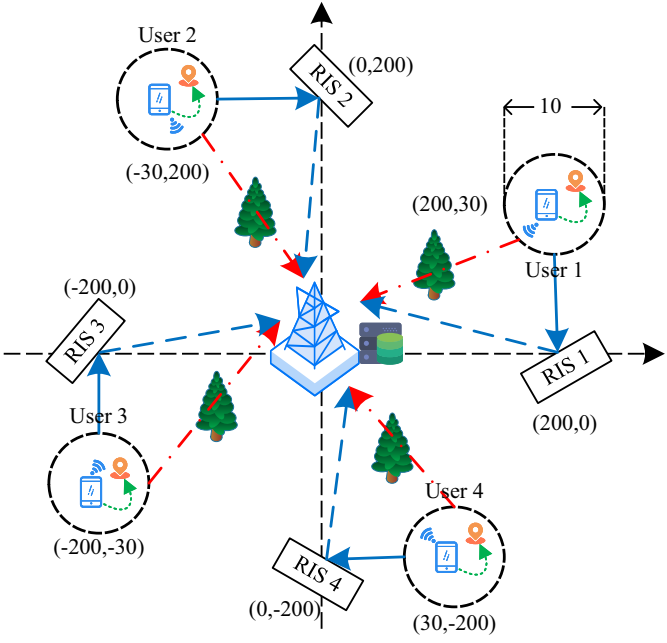}
  \caption{The simulation deployment of the RIS-aided BS-centric SR system.}
  \label{fig_2}
\end{figure}
In this section, numerical simulation results of MATLAB are provided to evaluate the performance of the proposed algorithm. A two-dimensional Euclidean coordinate system is used to illustrate the locations of BS, RISs, and users. Similar to \cite{16}, an RIS-based femtocell model is adopted, all measured in meters as shown in Fig. 2. Specifically, the location of BS is $\left( 0,0 \right)$ meters, we consider $K=4$ and the locations of RISs are $\left( 200,0 \right)$ meters, $\left( 0,200 \right)$ meters, $\left( -200,0 \right)$ meters, and $\left( 0,-200 \right)$ meters, respectively. Each user is randomly distributed in circles with a radius of 10 m at the center of the circle $\left( 200,30 \right)$ meters, $\left( -30,200 \right)$ meters, $\left( -200,-30 \right)$ meters, and $\left( 30,-200 \right)$ meters, respectively. Additionally, all RISs are equipped with same reflecting elements, while the BS is equipped with 4 antennas $\left( M=4 \right)$. Similar to \cite{3}, we set $Q=128$. The total bandwidth of system is 100 KHz. The noise power ${{\sigma }^{2}}$ is -100 dBm. The cycle of system $T$ is 5 s. The sensing parameters are given as $v_{k}^{s}=0.05$ Mbits/s, $v_{k}^{p}=2.5$ Mbits/s, and $p_{k}^{s}=0.5$ J/Mbits. The computing parameters are given as ${{k}_{l}}={{10}^{-25}}$ and $C=600$ cycles/bit.

We assume that all channels are adopted the Rician fading channel model as in \cite{8}, and thus ${\mathbf{h}_{\text{r},k}}$, ${\mathbf{G}_k}$, and ${\mathbf{h}_{\text{d},k}}$ can be given as ${{\mathbf{h}}_{\text{r},k}}={\sqrt{{L}_{\text{r},k}}}\left( \sqrt{\frac{K_{1}^{{}}}{K_{1}^{{}}+1}}\mathbf{h}_{\text{r},k}^{\text{LoS}}+\sqrt{\frac{1}{K_{1}^{{}}+1}}\mathbf{h}_{\text{r},k}^{\text{NLoS}} \right)$, ${{\mathbf{G}}_{k}}={\sqrt{{L}_{k}}}\left( \sqrt{\frac{K_{2}^{{}}}{K_{2}^{{}}+1}}\mathbf{G}_{k}^{\text{LoS}}+\sqrt{\frac{1}{K_{2}^{{}}+1}}\mathbf{G}_{k}^{\text{NLoS}} \right)$, and  ${{\mathbf{h}}_{\text{d},k}}={\sqrt{{L}_{\text{d},k}}}\mathbf{h}_{\text{d},k}^{\text{NLoS}}$, respectively. ${L_{{\rm{r}},k}}$, ${L_k}$, and ${L_{{\rm{d}},k}}$ denote the corresponding channel path loss and are modeled as ${{\beta }_{0}}{{\left( d \right)}^{{{\alpha }_{o}}}}$. ${{\bf{h}}_{{\rm{r}},k}^{{\rm{LoS}}}}$ and ${{\bf{G}}_k^{{\rm{LoS}}}}$ denote the corresponding steering vectors, and the components of ${{\bf{h}}_{{\rm{r}},k}^{{\rm{NLoS}}}}$, ${{\bf{G}}_k^{{\rm{NLoS}}}}$, and ${{\bf{h}}_{{\rm{d}},k}^{{\rm{NLoS}}}}$ follow the the distribution of $\mathcal{C}\mathcal{N}\left( 0,1 \right)$. In addition, we set ${{\beta }_{0}}=-30$ dB, and ${{K}_{1}}={{K}_{2}}=10$. The channel attenuation parameters ${{\alpha }_{0}}$ of user-BS, user-RIS, and RIS-BS are set as -2.0, -2.2 and -3.6, respectively.

\begin{figure}[!t]
  \centering
  \includegraphics[width=3.5in]{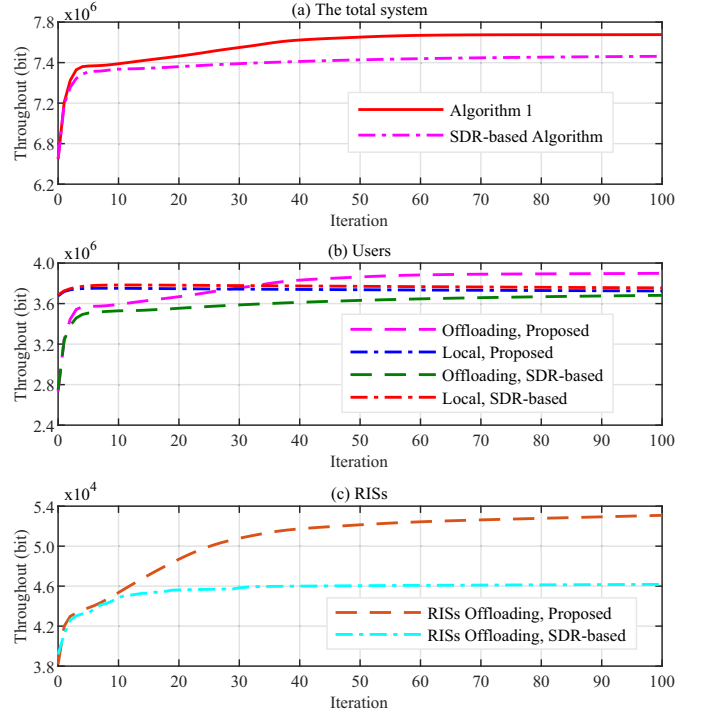}
  \caption{The convergence performance, where $E_{k}^{\max }=10$ J, $N=100$ and $\alpha =0.4$.}
  \label{fig_3}
\end{figure}
In order to highlight the effectiveness of the proposed algorithm and model, we set up five benchmark schemes:
\begin{itemize}
  \item \textbf{Without SR}: In this case, the RIS is only to improve the quality of user communication.
  \item \textbf{Random Phase Shift}: In this case, the phase shifts for each RIS are randomly generated at the beginning of optimization.
  \item \textbf{Without RIS}: In this scheme, the wireless channel gains via RIS are 0. Additionally, the optimal beamforming obtained in Section III-B is replaced by $\mathbf{W}=\left[ {\mathbf{w}_{1}},{\mathbf{w}_{2}},\cdots ,{\mathbf{w}_{K}} \right]=\mathbf{H}{{\left( {{\mathbf{H}}^{\text{H}}}\mathbf{H} \right)}^{-1}}$ via Zero Forcing method ($M\ge K$), where $\mathbf{H}=\left[ {\mathbf{h}_{\text{d},1}},{\mathbf{h}_{\text{d},2}},\cdots ,{\mathbf{h}_{\text{d},K}} \right]\in {{\mathbb{C}}^{M\times K}}$.
  \item \textbf{Local Computation}: In this case, all the energy of each user is used for local computation, which means that the energy allocated by users to offload computation is 0 J and there is no SR system.
  \item \textbf{Random Energy Partition $\boldsymbol{\beta}$}: In this case, the energy partitions for each user are randomly generated at the beginning of optimization.
\end{itemize}

Fig. 3 plots the convergence process of user offloading, local computation, and RIS offloading with the increasing of iteration steps, respectively when $E_{k}^{\max }=10$ J, $N=100$ and $\alpha =0.4$. First, we can observe that throughput in all figures gradually stabilizes with the increasing of iterations, and reaches convergence after around 30 to 50 according to Fig. 3(a). Second, according to Fig. 3(b), the users' throughput of offloading computation exceeds the throughput of local computation at step 30 under the proposed algorithm. This is because the channel-related variables gradually approach the optimal value with the iterative optimization. Besides, the performance of proposed algorithm outperforms the SDR-based algorithm, which demonstrates the effectiveness of our proposed algorithm. Furthermore, by combining with Fig. 3(b) and Fig. 3(c), we can see that RISs’ throughput increases with the increasing of users' offloading throughput. This is due to the fact that when the users' offloading throughput increases, the signals reflected by the RIS also increase, which makes the rate of RISs gradually increase under the SR-based system. This also verifies that RISs can act as IoT devices to enable IoT transmission without any dedicated RF components.

\begin{figure}[!t]
  \centering
  \includegraphics[width=3.5in]{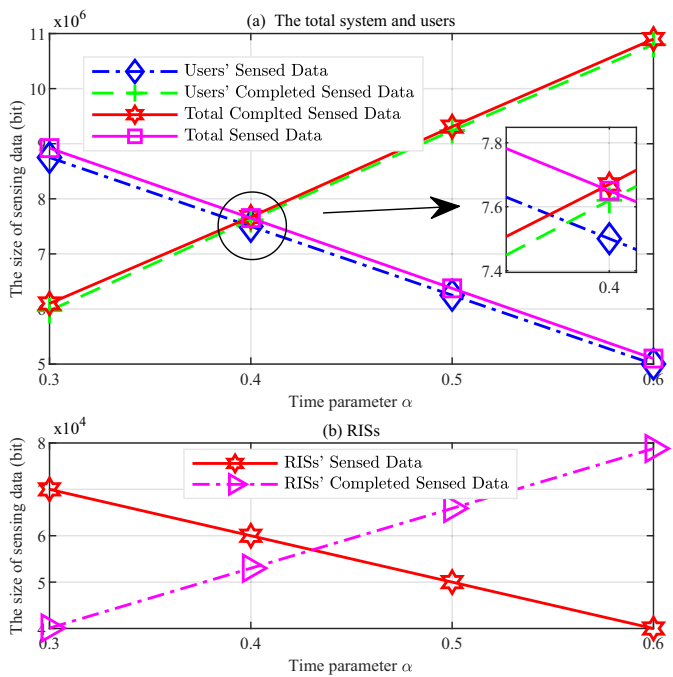}
  \caption{The impact of sensing constraints and time allocation parameter $\alpha$ on system performance, where ${{E}^{\max }}=10$ J and $N=100$.}
  \label{fig_4}
\end{figure}
Fig. 4 illustrates the impact of time allocation parameter $\alpha $ and sensing constraints on system performance when $E_{k}^{\max }=10$ J and $N=100$. Specifically, by setting different parameters $\alpha $, we compare the amount of data sensed by the IoT devices (users and RISs) in the first stage with the amount of data tackled by the IoT devices in the second stage to find a reasonable time allocation parameter. From these two figures, we can observe that the size of sensed data decreases with the time allocation coefficient increasing, and the amount of completed sensed data increases with the time allocation coefficient increasing, which means more time resources lead to more system performance, and a reasonable parameter $\alpha $ helps to balance sensing and offloading. More importantly, by considering constraint $\text{C}4$, it is easy to conclude that the performance of the system is best when the data sensed by IoT devices happens to be tackled, which means that the inequalities in constraint $\text{C}4$ are equations. Based on this conclusion and Fig. 4, we can conclude that the best parameter ranges of users, RISs, and total users and RISs are less than or equal to 0.39, 0.42, and 0.4, respectively. As a result, the optimal system parameter range in this simulation experiment is less than or equal to 0.39. Moreover, when the sensing rate of IoT devices increases, the energy consumed by the data sensing will increase and the energy used for tackling sensed data will decrease accordingly, which leads to the growth of sensed data and the decline of total completed sensed data. This means that the line representing sensing data will go up, the line representing completed sensed data will go down, and the best time allocation parameter gradually increases according to the trend in Fig. 4. In sum, we can conclude that the optimal time allocation parameter is related to the sensing capability of IoT devices.

\begin{figure}[!t]
  \centering
  \includegraphics[width=3.5in]{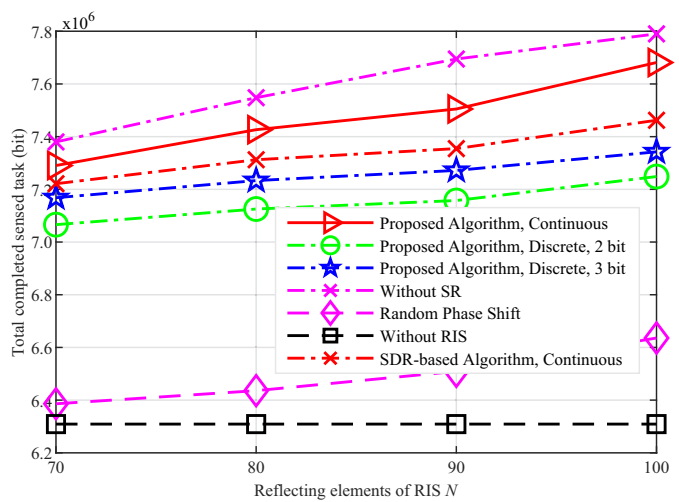}
  \caption{The total completed sensed data versus the number of reflecting elements $N$, where  ${{E}^{\max }}=10$ J and $\alpha=0.4$.}
  \label{fig_5}
\end{figure}
Fig. 5 compares the total completed sensed data under the proposed design and benchmark schemes with respect to the number of reflecting elements when ${{E}^{\max }}=10$ J and $\alpha=0.4$. Some observations can be made from the figure. First, with the growth of reflecting elements, the performance of all schemes simultaneously increases except for “Without RIS” scheme. This is because the quality of channels between users and BS becomes better with the aid of RIS. Second, we can see that “Without SR” scheme achieves higher performance compared to proposed algorithm. This is due to the backscatter channel ${\mathbf{b}_{k}}$ changes with ${{c}_{k}}\in \left\{ 0,1 \right\}$ and the effect of such changes on system performance may be significant, and the system needs to behave well for all situations, which leads to a stricter requirement for proposed algorithm. Third, compared to “Without RIS” and “Random Phase Shift” schemes, the proposed scheme can always obtain better performance, which indicates that the benefits of RIS, and the importance of optimization of $\mathbf{W}$ and $\boldsymbol{\Theta } $. Additionally, there is a performance gap between the “Proposed Algorithm” and “SDR-based Algorithm”. This is due to the fact that compared to the element iterative method, the complexity of SDR technique is high and there is a certain performance loss. Finally, compared to the ‘2-bit’ phase shift scheme, it is worth pointing out that the performance of the ‘3-bit’ phase shift scheme is closest to the performance of the continuous phase shift scheme.

\begin{figure}[!t]
  \centering
  \includegraphics[width=3.5in]{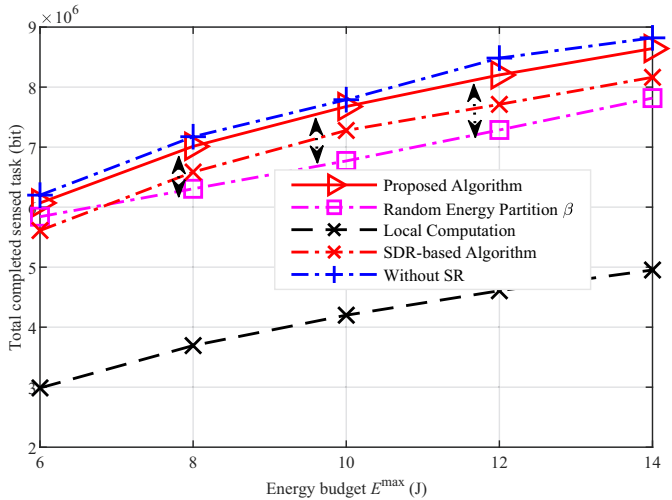}
  \caption{The total completed sensed data versus different energy budget $E_{}^{\max }$, where $\alpha =0.4$ and $N=100$.}
  \label{fig_6}
\end{figure}
Fig. 6 shows the total completed sensed data with respect to the users’ uniform energy budget when $\alpha =0.4$ and $N=100$. From the figure, some observations can be indicated. First, as expected the total completed sensed data of all schemes increases with the more energy budget. Second, the performance gap between proposed algorithm and random energy partition $\boldsymbol{\beta}$ gradually becomes lager with the increasing of energy budget. This is because the energy partition strategy of proposed algorithm is always optimized, and the increasing energy used by the user to offload can also indirectly increase the total completed sensed data under the gains of the SR system, which leads to additional throughput that is offloaded by the passive communication enabled by the RISs via the users’ incident signals. Moreover, compared to “Without SR” scheme, the proposed scheme is inferior in the total completed sensed data, but the gap is not very large, and the RISs can play an additional sensing role in the proposed scheme. Thus, we can conclude that more devices are allowed to sense environment by sacrificing some performance in exchange when the requirements of data performance are not very strict. In this case, the proposed scheme can behave better than other benchmarks. Finally, with the increasing of energy budget, there is still a performance gap between “SDR-based Algorithm” and “Proposed Algorithm”, which verifies the effectiveness of our proposed algorithm.

\begin{figure}[!t]
  \centering
  \includegraphics[width=3.5in]{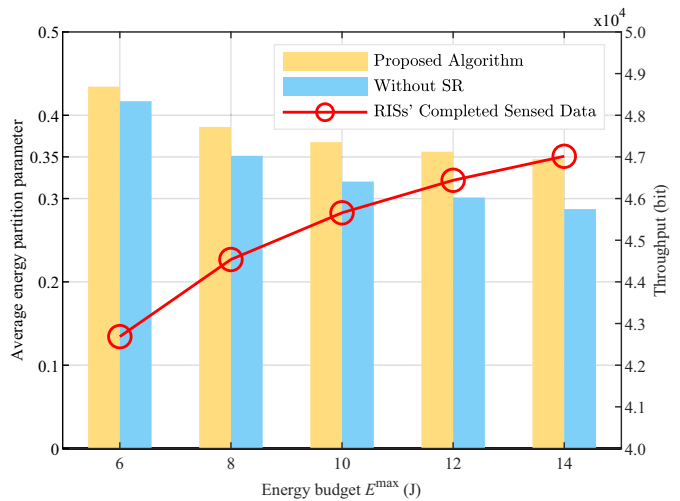}
  \caption{The SR performance versus different energy budget $E_{}^{\max }$, where $N=100$ and $\alpha=0.4$.}
  \label{fig_7}
\end{figure}
In order to further investigate the performance of SR-based system, Fig. 7 depicts the trend of average energy partition parameter and RISs’ completed sensed data with respect to different energy budgets, where $\alpha =0.4$ and $N=100$. As shown in the figure, the average energy partition parameter decreases  when $E_{}^{\max }$  is large. This is due to the fact that the users' transmitting power increases with the more energy budget, and the users’ offloading rate gradually approaches the limit of transmission rate. 
%As a result, the users’ energy consumed by the offloading is still a stable value during cycle $T$, even if the energy budget increases. 
Another observation is that the amount of sensed data tackled by RIS increases with $E_{}^{\max }$, and the increase of RISs’ completed sensed data gradually slows down. Two reasons explain these results. On one hand, although the average energy partition parameter decreases with $E_{}^{\max }$, the energy used by the user for offloading is actually increased. This means that the signals reflected via RIS are gradually increasing. On the other hand, when the users’ offloading rate gradually approaches the limit of transmission rate, the increase of signals reflected via RISs also gradually slow down. It is noted that under the same energy budget, the average energy partition parameter in SR system is higher than that of without SR scheme. This is because the backscatter channel ${\mathbf{b}_{k}}$ is affected by the RIS symbol ${{c}_{k}}$ in the SR system. Compared to the additional channel without SR, the backscatter channel will consume more energy. However, with the help of RISs in the SR system, i.e., RISs acting as IoT devices to sense and deal with data, more IoT devices can access the SR system for sensing.

\section{Conclusions}
In this paper, we investigated an RIS-aided BS-centric SR system, where the SR-based system and “\emph{sense-then-offload}” protocol are considered. We formulated the total completed sensed bits of all users and RISs maximization problem by jointly optimizing the time allocation parameter, the passive beamforming at each RIS, the transmit beamforming at BS, and the energy partition parameters for all users under the constraints on the size of sensed data and energy supply. To address the nonconvex problem, we proposed an AO-based algorithm where Monte Carlo, Lagrangian dual transformation and FP are applied to convert the original optimization problem into a solvable one. Simulation results verified that the RIS-aided SR system outperforms other benchmarks in sensing. Besides, the channel and system performance can be significantly improved with the aid of RIS. In our future work, we will further investigate data sensing and computation offloading in multi-BS collaboration scenario.

{\appendix[Proof of the Proposition 1]
As shown in Eq. (\ref{eq31}) and Eq. (\ref{eq32}), it is evident that ${{{A}}_{1k}}$ and ${{{A}}_{2k}}$ are greater than 0. In addition, by combining Eq. (\ref{eq29}), Eq. (\ref{eq33}) and Eq. (\ref{eq34}), it is easy to show that ${{{B}}_{1k}}\ge 0$ and ${{{B}}_{2k}}\ge 0$. For ease of expression, according to Eq. (\ref{eq9}) and Eq. (\ref{eq30}), we simplified $ f_{4{\beta _k}}^{}\left( {{\beta _k}} \right)$ into the following function
\begin{align}
  \label{eq35}
  f\left( x \right)=-ax+b\sqrt{x}+c\sqrt{1-x},
\end{align}
where $a = {\left( {{{{A}}_{1k}} + {{{A}}_{2k}}} \right)}\alpha T{(E_{k}^{\text{o}})}^{-1} > 0$, $b = {\left( {{{{B}}_{1k}} + {{{B}}_{2k}}} \right)}\sqrt{\alpha T{(E_{k}^{\text{o}})}^{-1}} > 0$, $c = {\alpha T}{(C)_{}}^{-1}\sqrt[3]{{E_{k}^{\text{o}}}{(\alpha T{{\kappa }_{l}})^{-1}}}>0$, and $x = {{\beta }_{k}}\in \left[ 0,1 \right]$. And the corresponding second-order derivative can be given by
\begin{align}
  \label{eq36}
  \frac{{{\partial }^{2}}f\left( x \right)}{\partial {{x}^{2}}}=-\frac{b}{4}{{x}^{-\frac{3}{2}}}-\frac{c}{4}{{\left( 1-x \right)}^{-\frac{3}{2}}}<0.
\end{align}
Thus, the objective function of Problem (\ref{p8}) is concave with respect to ${\beta _k}$.
}

\bibliographystyle{IEEEtran.bst}
\bibliography{myref}

\end{document}